\documentclass[aps,prapplied,reprint,amsmath,amssymb,floatfix]{revtex4-2}

\usepackage{graphicx}
\usepackage{bm}
\usepackage{mathtools}
\usepackage[
  colorlinks=true,
  linkcolor=blue,
  citecolor=blue,
  urlcolor=blue
]{hyperref}
\usepackage{subcaption}
\graphicspath{{figures/}}

\begin{document}





\title{Pound--Drever--Hall Feedforward for Trapped-Ion Optical Qubits}

\author{Jia-Yang Gao}
\email{e0941514@u.nus.edu}
\affiliation{Centre for Quantum Technologies, National University of Singapore,
Singapore 117543, Singapore}
\author{Jianwei Lee}
\affiliation{Centre for Quantum Technologies, National University of Singapore,
Singapore 117543, Singapore}
\author{Morteza Ahmadi}
\affiliation{Centre for Quantum Technologies, National University of Singapore,
Singapore 117543, Singapore}
\author{Manas Mukherjee}
\affiliation{Centre for Quantum Technologies, National University of Singapore,
Singapore 117543, Singapore}

\date{\today}

\begin{abstract} 

Laser phase noise is one of the limiting factors that dictate the gate fidelities and coherence time in trapped-ion quantum systems.
Previously, it has been reported that when the Rabi frequency of the ion qubit is close to the servo bump phase-noise frequency, the driving laser limits the fidelity and coherence time. This issue has typically been mitigated by choosing a Rabi frequency outside the servo-bump region. However, this constrains the usable range of gate speeds and can limit the achievable fidelity.
To address this issue, we developed an active phase-noise stabilization for the barium ion optical qubit laser at 1762 nm, employing a fiber electro-optic modulator (EOM) with an electrical feedforward servo. 
Our results demonstrate that this setup based on Pound–Drever–Hall (PDH) feedforward method can suppress 15 dB servo bump phase noise near the bump peak frequency in our locking system. The laser phase noise is analyzed using delayed self-heterodyne interferometry (DSHI). We further tested the stabilized laser on the optical qubit, observing a clear improvement in coherence time based on the measured amplitude decay of Rabi oscillations even when the Rabi frequency lies within the servo bump bandwidth. This technique can be readily adapted to other optical qubits with minimal modifications.



\end{abstract}

\maketitle

\section{Introduction}

In recent years, quantum computing has emerged as a promising field with diverse potential applications across multiple domains, including machine learning~\cite{chen2020variational,mitarai2018quantum,nguyen2021quantum,dutta2022single}, finance~\cite{lim2024quantum}, and sensing~\cite{dutta2020single}. Among various quantum computing platforms, the trapped-ion system stands out as a leading candidate~\cite{cirac1995quantum,monroe1995demonstration} due to its long qubit coherence times and high-fidelity gate operations~\cite{wang2021single}. Single- and two-qubit gates in trapped-ion systems can be implemented using optical, microwave, or radio-frequency fields. For laser-driven optical-qubit gates, phase and intensity fluctuations in the control laser can degrade the gate fidelity~\cite{day2022limits,jiang2023sensitivity,nakav2023effect}. In particular, the influence of laser phase noise is determined by the spectral overlap between the phase-noise power spectral density (PSD) and the filter function of the driven qubit ~\cite{jiang2023sensitivity,nakav2023effect,day2022limits,soare2014experimental}. Low-frequency fluctuations may produce slowly varying detuning and phase errors, whereas fast noise near the characteristic gate-response frequencies can directly reduce the fidelity of single- and two-qubit operations. Suppressing both phase and intensity noise is therefore essential for high-fidelity quantum control and fault-tolerant operation. When a laser is locked to a high-finesse reference cavity, the feedback controller effectively suppresses low-frequency phase and frequency noise. However, due to its finite bandwidth and unity-gain roll-off, it inevitably amplifies noise near the loop bandwidth, leading to the well-known servo bump—a pair of peaks in the frequency-noise PSD symmetrically displaced from the carrier. These features are often modeled as Gaussian-shaped bumps superimposed on a white-noise floor and typically appear near the servo’s unity-gain frequency (on the order of $10^5$–$10^6$ Hz, depending on the actuator).

In practice, servo-bump phase noise can be addressed through different approaches:

\textbf{(i) Spectral filtering:} Passing the locked laser through a high-finesse ``cleanup'' cavity suppresses phase noise at Fourier frequencies above the cavity linewidth—the cavity transfer function \(T(\omega)\) acts as a low-pass filter—so the servo bumps are strongly attenuated in the transmitted light \cite{hald2005efficient}. In power-hungry applications, the filtered output is then amplified downstream with a low-noise stage. Some implementations explicitly use cavity filtering to tame MHz-scale bumps relevant to fast quantum control~\cite{krinner2024low,semenin2025improved,akerman2015universal}. 

\textbf{(ii) Reducing free-running noise or extending feedback bandwidth.} Designing low-noise lasers and optimizing servo architectures jointly mitigate servo-bump effects. Long external-cavity diode lasers (ECDLs) lower the intrinsic free-running phase-noise floor—e.g., a 20-cm cavity can reduce noise by nearly two orders of magnitude compared to short-cavity designs~\cite{avalos2023field}—allowing stabilization with smaller loop gain and reduced bump height. Extending the feedback bandwidth can also be done by introducing an intra-cavity EOM into the ECDL \cite{le2009wide}.



\textbf{(iii) Feedforward methods.}
In addition to feedback, various feedforward schemes have been explored to cancel servo-bump noise beyond feedback’s bandwidth limits \cite{bagheri2009semiconductor,aflatouni2012wideband,watts2016phase,lintz2017note,chen2019feedforward,scharnhorst2015high,zhang2016linewidth,huang2017precise,li2019laser,parniak2021high,li2017efficient}. For example, Li~\textit{et~al.}~\cite{li2022active} demonstrated a feedforward implementations with heterodyne measurements between the laser field and its transmission through a filtering cavity to monitor high-frequency phase noise. However, this requires an additional feedback loop to compensate for drifts in the optical path length between the two arms of the heterodyne setup. Since the PDH signal inherently encodes the full spectral content of the laser’s phase noise, it is natural to ask whether the PDH technique—typically used only for feedback—can also be leveraged for feedforward control. A particularly powerful and increasingly adopted approach is \textbf{PDH feedforward} ~\cite{chao2024pound}. In this method, the residual PDH error signal (which encodes the laser’s instantaneous phase fluctuations after locking) is recycled and applied to a fast phase actuator (e.g., an EOM) with a matched delay to cancel the bump-induced phase noise. This technique can suppress bump noise power by \(\sim 40\,\mathrm{dB}\) around a few megahertz.

Recent work has demonstrated growing interest in deploying the PDH feedforward method across diverse applications. For example, Maddox \emph{et al.}~\cite{maddox2024enhanced} used this technique to suppress laser phase noise in stimulated Raman adiabatic passage (STIRAP) state transfer of ultracold RbCs molecules. For the applicability of PDH feedforward stabilization to optical qubit operations —particularly lasers driving quadrupole transitions in trapped-ion systems—has not been experimentally verified. As optical qubits are based on ground and metastable state coupling, the required laser characteristics are also stringent in terms of linewidth, random phase mixing, spectral purity etc. Therefore, it is not a priori to assume that PDH feedforward would work for optical qubits. 

In this work, we experimentally demonstrate that the PDH feedforward method can be effectively applied to a laser used for an optical qubit gate implementation. We further demonstrate the effectiveness of the method by operating the qubit with a Rabi frequency near the servo-bump peak, while substantially reducing the associated coherence degradation. This demonstration will improve the frequency operating range of an qubit, an important criteria to avoid frequency crowding while scaling the number of qubits. 

We developed a phase-noise stabilization setup based on PDH feedforward method for a 1762~nm laser driving the $\mathrm{S}_{1/2} \leftrightarrow \mathrm{D}_{5/2}$ quadrupole transition in ${}^{138}\mathrm{Ba}^{+}$, employing a fiber electro-optic modulator (fiber EOM) with an electrical feedforward servo. Our results show that this setup suppresses the 15dB servo-bump phase noise created by the PDH locking system near the bump peak frequency, without broadening the laser linewidth. The phase-noise and linewidth characteristics are analyzed using delayed self-heterodyne interferometry (DSHI). We validate the impact on qubit performance by testing the stabilized beam with a Doppler-cooled single ion: from Rabi oscillations driven near the servo-bump frequency, we observe an improvement in coherence time. These results provide the first experimental verification that PDH feedforward can enhance coherence in optical qubit gates, confirming its suitability for trapped-ion quantum processors.




This article is structured as follows. We first describe the working principle of the PDH feedforward method and present the experimental phase-noise stabilization setup used for the 1762 nm optical-qubit laser. We then introduce delayed self-heterodyne interferometry (DSHI), describe the DSHI implementation used to characterize the laser phase noise and linewidth, and present the measured suppression of the servo-bump phase noise together with the linewidth comparison.

To connect the optical phase-noise measurements to qubit performance, we next develop a finite-time response model for a resonantly driven optical qubit subject to stochastic detuning noise. Starting from the toggling-frame stochastic differential equation, we derive the corresponding finite-time dynamical map, express the result in terms of a Pauli-transfer matrix, and relate this matrix response to the phase-noise-limited average gate infidelity. We also introduce the scalar Rabi-envelope approximation, which shows how resonant Rabi oscillations filter laser phase noise near the Rabi frequency.

We then validate the stabilized laser in a trapped-ion experiment using a single \(^{138}\mathrm{Ba}^{+}\) optical qubit. After describing the ion-trap setup and measurement sequence, we present Rabi-oscillation measurements performed with the Rabi frequency close to the servo-bump frequency, showing a clear improvement in coherence time and \(\pi\)-pulse state-transfer performance when the PDH feedforward stabilization is enabled. We also present a frequency scan of the 1762 nm laser that reveals off-resonant excitation associated with the servo-bump phase noise. Finally, we quantify the coherence-time improvement as a function of servo-bump phase-noise suppression, discuss the relevance of this technique for optical-qubit operations and future two-qubit gates, and conclude with a summary of the main results.

\section{\label{sec: Methods} Methods}

\begin{figure*}[t!]
  \centering
  \includegraphics[width=1.0\linewidth]{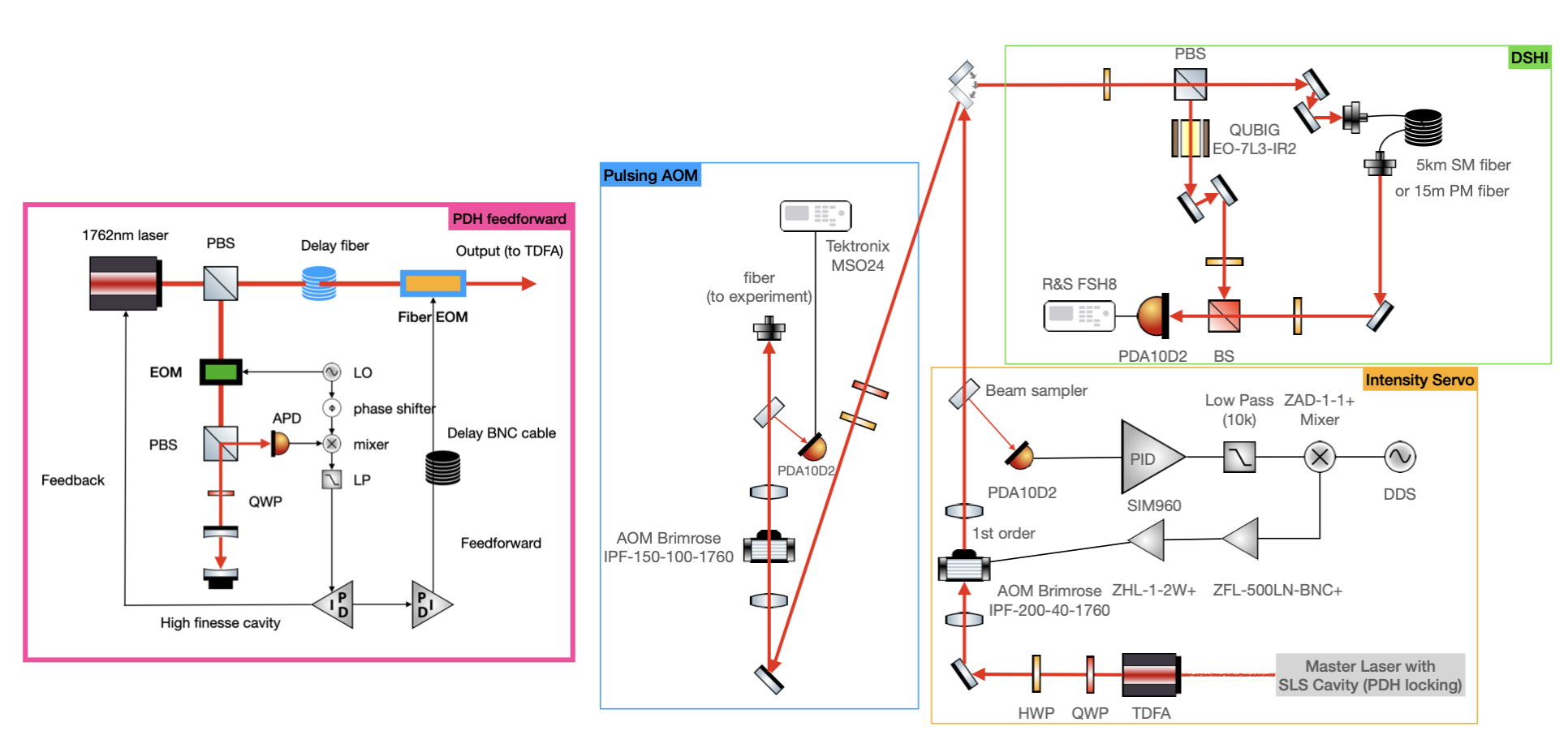}
  \caption[Experimental setup]{
  {\bfseries Experimental setup for phase noise stabilization, DSHI measurement, intensity servo, and pulsing AOM} 
  The PDH feedforward setup recycles the PID control signal through a delayed BNC cable, which is then applied as feedforward to a fiber electro-optic modulator (Jenoptik PM1750). 
  A delayed fiber is used to match the optical phase between the two arms, and the output serves as the seed laser for the thulium-doped fiber amplifier (TDFA). 
  For intensity stabilization, the signal from a photodetector is used for electrical feedback to control the RF drive of the acousto-optic modulator (AOM). 
  In the delayed self-heterodyne interferometer (DSHI) measurement, one path is delayed by the optical fiber, and the EOM introduces a frequency shift for heterodyne detection. 
  The pulsing AOM (Brimrose IPF-150-100-1760) is driven by a system comprising a field-programmable gate array (FPGA) and a direct digital synthesizer (DDS), generating the pulse sequences used in the ion trap experiment.
  }
  \label{fig:PDHFF}
\end{figure*}

\subsection{PDH Feedforward Working Principle}

The PDH feedforward scheme builds upon a laser that is already tightly locked to an optical cavity through conventional PDH feedback control. Under this condition, the residual phase noise $\phi(t)$ of the locked laser is small, and its instantaneous linewidth is typically much narrower than the cavity linewidth $\gamma_c$. The remaining noise manifests as servo bumps at characteristic angular frequencies $\omega_n \gg \gamma_c$, beyond the bandwidth of the feedback loop.

The optical field incident on the cavity can thus be expressed as
\begin{equation}
E_{\mathrm{in}}(t) = E_0 \exp\{i[\omega_c t + \phi(t)]\},
\end{equation}
where $E_0$ is the real amplitude, $\omega_c$ the cavity resonance, and $\phi(t)$ the residual phase fluctuation. Phase modulation at frequency $\Omega$ with modulation depth $\beta$ is applied through an electro-optic modulator (EOM), producing
\begin{equation}
E_{\mathrm{in}}(t) = E_0 \sum_{k=-\infty}^{\infty} J_k(\beta)
\, \exp\{i[(\omega_c + k\Omega)t + \phi(t)]\},
\end{equation}
where $J_k(\beta)$ is the $k$-th order Bessel function. Since the modulation frequency $\Omega$ and the servo bump frequencies $\omega_n$ are much larger than the cavity linewidth $\gamma_c$, only the carrier at $\omega_c$ couples strongly into the cavity; the sidebands are effectively reflected.

Assuming the on-resonance carrier is fully transmitted and the sidebands completely reflected, the reflected field is
\begin{equation}
E_{\mathrm{ref}}(t) = -E_{\mathrm{in}}(t) + E_0 J_0(\beta) e^{i\omega_c t},
\end{equation}
where the negative sign represents the $\pi$ phase shift on reflection. After demodulation and low-pass filtering, the PDH discriminator produces an error signal proportional to the sine of the instantaneous phase \cite{chao2024pound}:
\begin{equation}
V_{\mathrm{err}}(t) \propto \sin[\phi(t)] \approx \phi(t).
\end{equation}
This residual error therefore encodes the instantaneous phase noise of the laser even at Fourier frequencies beyond the feedback bandwidth.

To cancel this noise, the error signal is amplified with constant proportional gain $G_{\mathrm{ff}}$ and applied, after an appropriate delay $\tau$, to a second EOM that directly modulates the laser output phase. The output phase is then
\begin{equation}
\phi_{\mathrm{out}}(t) = \phi(t) + G_{\mathrm{ff}} \sin[\phi(t-\tau)] \approx \phi(t) + G_{\mathrm{ff}} \phi(t-\tau).
\end{equation}
When the residual fluctuations are small, choosing $G_{\mathrm{ff}} = -1$ and matching the delay ($t' = t$) yields nearly complete cancellation of $\phi(t)$.

In practice, the delay $\tau$ compensates for the propagation time of the optical and electronic paths from the laser through the cavity and PDH detection to the feedforward EOM. A fiber or coaxial line is inserted so that $2\pi f_b \tau = 2\pi n$ at the dominant servo-bump frequency $f_b$, ensuring complete cancellation. This approach transforms the residual PDH signal into a real-time phase correction, achieving suppression comparable to that of cavity filtering but without limited by the transmission optical power through the cavity.

\subsection{\label{sec: phase noise stabilization setup} Phase noise stabilization setup}


The PDH feedforward servo setup shown in \figureautorefname{\ref{fig:PDHFF}} consists of an external-cavity diode laser (ECDL) at a 1762 nm wavelength. The linewidth is further reduced using the standard Pound–Drever–Hall (PDH) technique locked to a Ultra Low Expansion (ULE) cavity using 5 MHz sideband. This ULE cavity, supplied by Stable Laser Systems, features a horizontal design and provides a frequency stability of approximately $100~$Hz. The measured cavity linewidth of 1 kHz is consistent with the manufacturer’s specifications. By recycling the PID control signal through a $325 ~$m RG-58 BNC delay cable, an inverted-gain setting in the FALC, and an RF amplifier (ZHL-32A+), the signal is applied as feedforward to a fiber electro-optic modulator (Jenoptik PM1750). This configuration effectively suppresses the servo-bump phase noise originating from the PDH locking loop. A $20~$m PM1550 delay fiber is inserted to compensate for the phase mismatch between the two arms. The output beam is subsequently amplified using Thulium-doped fiber amplifiers (TDFAs), which provide sufficient optical power for implementing optical qubit gate operations in the ion-trap quantum processor.

Since the servo-bump peak in our system appears at approximately 132 kHz dependent of the gain setting, which is significantly lower than the MHz-range bumps reported in other works, we adjusted the BNC cable length to achieve optimal feedforward delay at this frequency. We first estimated the cable length required to introduce a $180°$ phase shift at $132~$kHz, then fine-tuned it experimentally by testing nearby lengths. The optimal length was determined from the DSHI phase-noise spectrum, where a flattened bump region indicated proper delay matching.

\begin{figure*}[ht!]
    \centering
    \begin{subfigure}[b]{0.32\linewidth}
        \includegraphics[width=\linewidth]{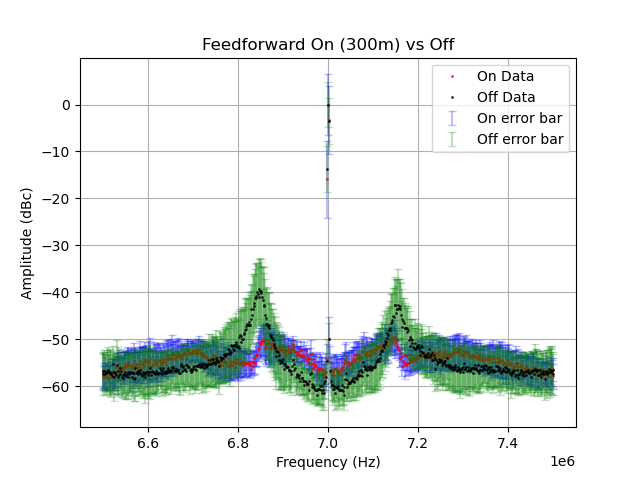}
        \caption{DSHI measurement with 300~m BNC cable.}
        \label{fig:DSHI_300}
    \end{subfigure}
    \hfill
    \begin{subfigure}[b]{0.32\linewidth}
        \includegraphics[width=\linewidth]{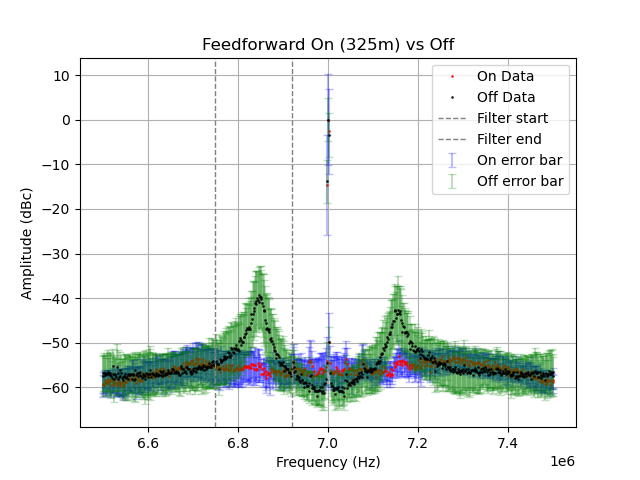}
        \caption{DSHI measurement with 325~m BNC cable.}
        \label{fig:DSHI_325}
    \end{subfigure}
    \hfill
    \begin{subfigure}[b]{0.32\linewidth}
        \includegraphics[width=\linewidth]{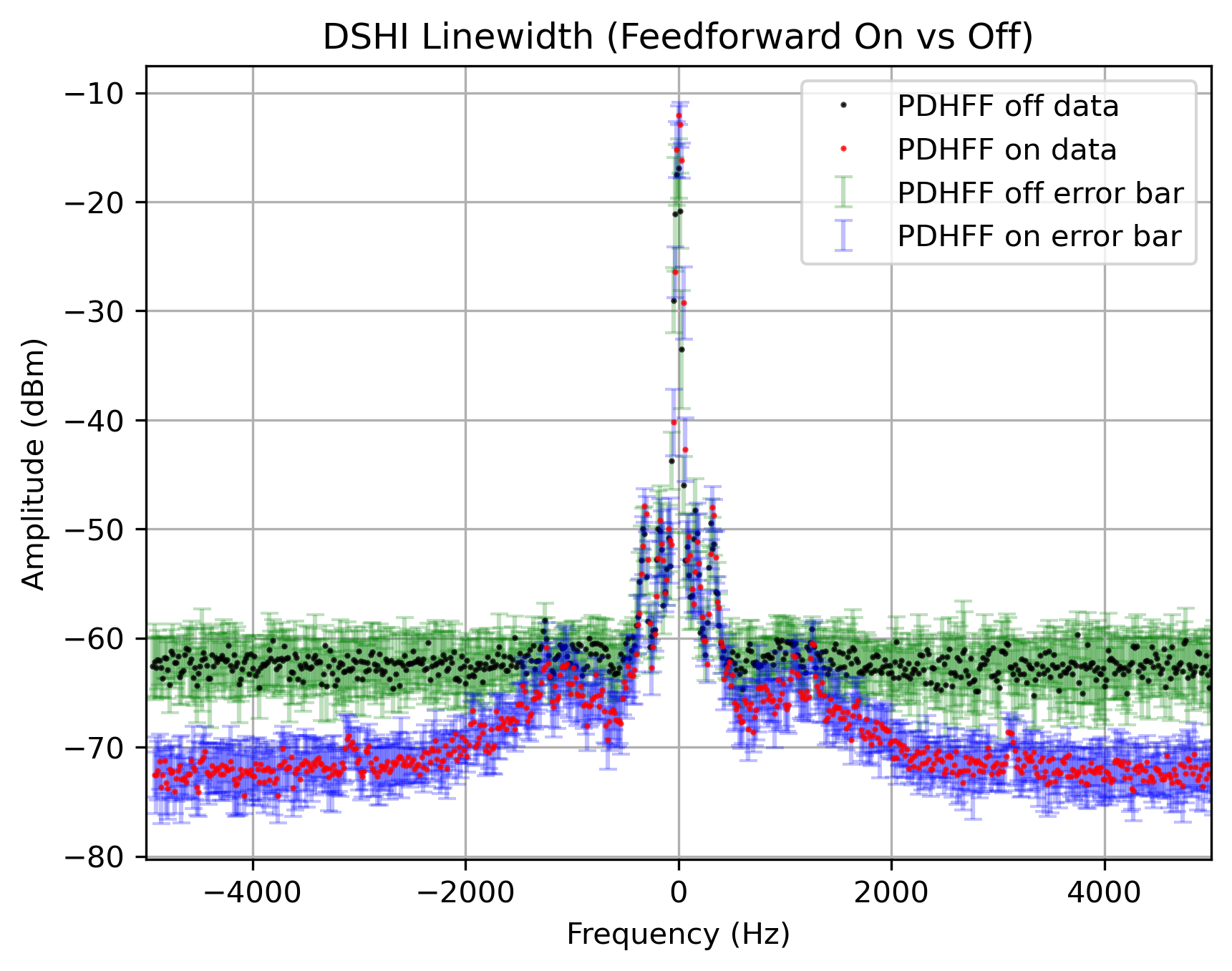}
        \caption{Extracted linewidth comparison.}
        \label{fig:DSHI_linewidth}
    \end{subfigure}

    \caption[
        DSHI phase-noise and linewidth measurements
    ]{
        \textbf{DSHI measurements of the phase noise and linewidth of the 1762~nm laser.}
        \autoref{fig:DSHI_300} shows that when the BNC cable length is not optimal (300~m), the phase noise cannot be fully reduced.
        \autoref{fig:DSHI_325} demonstrates that at the optimal cable length (325~m), the feedforward servo suppresses the servo-bump phase noise by up to 15~dB.
        \autoref{fig:DSHI_linewidth} shows that the feedforward servo does not broaden the laser linewidth, ensuring minimal impact on the qubit transition.
    }
    \label{fig:DSHI_phase_noise}
\end{figure*}

\subsection{Principle of Delayed Self-Heterodyne Interferometry}

The delayed self-heterodyne interferometry (DSHI) converts optical phase fluctuations into a measurable RF spectrum by interfering two copies of the same laser field separated by a delay~$\tau$ \cite{okoshi1980novel}.  In our implementation, the laser output is split into two arms: one passes through an electro-optic modulator (EOM) driven at frequency~$f_m$, while the other traverses a fiber delay line introducing a delay~$\tau$. When the two beams are recombined on a photodetector, the interfering photocurrent is
\begin{equation}
i(t) \propto \cos\!\big(2\pi f_m t + \omega_0 \tau + \phi(t) - \phi(t-\tau)\big),
\end{equation}
where $\phi(t)$ denotes the laser phase and $\omega_0$ the optical carrier. The RF tone at $f_m$ therefore carries an instantaneous phase
\begin{equation}
\Phi(t) = 2\pi f_m t + \Phi_0 + \Delta\phi(t),
\qquad
\Delta\phi(t) \equiv \phi(t) - \phi(t-\tau),
\end{equation}
so the random part of the beat phase equals the phase difference between two time-separated copies of the laser field.

For a stationary Gaussian phase process with single-sided power spectral density (PSD)~$S_{\phi}(f)$, the Fourier transform property of a delay gives
\begin{equation}
\Delta\Phi(f)
= \big(1 - e^{-i2\pi f\tau}\big)\,\Phi(f),
\end{equation}
leading to the PSD of the difference process
\begin{equation}
S_{\Delta\phi}(f)
= \big|1 - e^{-i2\pi f\tau}\big|^2 S_{\phi}(f)
= 4 \sin^2(\pi f \tau)\, S_{\phi}(f).
\end{equation}
Hence the measured RF noise pedestal around $f_m$ is the true phase-noise PSD weighted by the DSHI transfer function $H_{\tau}(f) = 4\sin^2(\pi f\tau)$. The interferometer thus acts as a frequency-dependent filter mapping optical phase noise to the RF domain.

We discuss different regimes and its characteristic:

\smallskip
\noindent\textbf{Short-delay regime.}
This regime satisfies
\[
    \Delta\nu \ll 1/\tau ,
\]
so the two arms remain mutually coherent. If the analyzer span is also smaller than
the fringe spacing, $\mathrm{Span}\ll 1/\tau$, the fringes are not resolved. The RF
spectrum around $f_m$ then appears as a smooth pedestal proportional to
\[
    H_\tau(f)=4\sin^2(\pi f\tau)
\]
up to frequencies of order $1/\tau$. In our phase-noise measurements, the
$15~$m fiber gives $\tau\simeq75$~ns and $1/\tau\simeq13$~MHz. With
RBW = 1~kHz and Span = 1~MHz, the condition $\mathrm{Span}\ll1/\tau$ is
satisfied. Therefore no fringes are visible, and the measurement directly maps
high-frequency phase noise, including servo bumps.

\smallskip
\noindent\textbf{Finite-delay regime.}
This regime also satisfies
\[
    \Delta\nu \ll 1/\tau ,
\]
but the analyzer span and resolution can make the finite-delay fringes visible.
The RF beat has fringe spacing $1/\tau$. Fringes are resolved when
$\mathrm{RBW}\ll1/\tau$ and the span covers several fringe periods. This applies
to the $5~$km interferometer, for which $\tau\simeq24.5~\mu$s and
$1/\tau\simeq40.8$~kHz, when the analyzer span covers multiple fringe periods.

In our linewidth runs, however, RBW = 30~Hz and Span = 20~kHz, which is smaller
than $1/\tau$. The full fringe pattern is therefore not covered. We fit the
central part of the spectrum using a finite-delay self-heterodyne model, with
the delay $\tau$ fixed. This gives the Lorentzian contribution associated with
the intrinsic linewidth and the Gaussian contribution associated with slow drift
and technical noise.

\smallskip
\noindent\textbf{Long-delay incoherent regime.}
In the long-delay limit,
\[
    \Delta\nu \gg 1/\tau ,
\]
the mutual coherence between the two arms vanishes. The RF spectrum then becomes
the self-convolution of the optical lineshape. Since our laser linewidth is of
order 100~Hz, this regime would require an impractically long fiber and is not
used for linewidth extraction in this work.

By combining short-delay ($15~$m) and long-delay ($5~$km) DSHI measurements, we capture both the high-frequency phase-noise spectrum and the intrinsic linewidth: the $15~$m interferometer resolves servo-bump suppression, while the $5~$km interferometer determines the linewidth.

\subsection{\label{sec: delayed self-heterodyne measurement} Delayed self-heterodyne measurement}

The first order beam output from the $200~$MHz AOM is split into two paths. In the first path, the laser travels through a resonant EOM (QUBIG, EO-7L3-IR2) and waveplates for polarization adjustment before reaching the photodetector (PDA10D2). In the second path, the laser is delayed by a $5~$km long single-mode telecom wavelength fiber if we measure the linewidth or a $15~$m short fiber if we want to characterize the servo bump phase noise only. The beam passes through waveplates for polarization adjustment as well. The polarization and spatial mode of the laser in both paths needs to be matched. The purpose of the long delay line is to partially destroy the coherence between the two paths, and its length is determined by the estimated coherence time of the laser under test \cite{ahmadi2024scalable}. The interference is created on the photodetector (PDA10D2) and monitored using an electronic spectrum analyzer (FSH8). Since this FSH8 provides a resolution down to $1~$Hz, this heterodyning technique allows us to measure the linewidth with high resolution. \figureautorefname{~\ref{fig:PDHFF}}  demonstrates the DSHI setup used in our experiment.
\figureautorefname{~\ref{fig:DSHI_phase_noise}} shows the phase noise and linewidth analysis for our $1762~$nm laser when using PDH feedforward method. As shown, the feedforward system stabilizes the $15~$dB servo bump phase noise from locking controller near bump peak frequency and does not broaden the laser's linewidth. Next, we will introduce the theoretical models to understand the impact of servo noise on an optical qubit.



\section{Trajectory model and toggling-frame SDE}

In order to understand the impact of servo noise on the optical qubit, we first develop the generic noise model. Consider a resonant optical qubit in the rotating frame of the mean laser frequency.  After removing deterministic detuning and calibration errors, residual laser phase noise enters as stochastic detuning noise,
\begin{equation}
H(t)=\frac{\hbar}{2}\left[\Omega\sigma_x+\beta(t)\sigma_z\right],\qquad \beta(t)=\dot\phi(t),
\label{eq:H}
\end{equation}
where $\Omega$ is the mean Rabi frequency and $\phi(t)$ is the residual optical phase in the qubit frame.  The stochastic state vector $\tilde{\bm c}$ obeys
\begin{equation}
\frac{d\tilde{\bm c}}{dt}=-\frac{i}{2}\left[\Omega\sigma_x+\beta(t)\sigma_z\right]\tilde{\bm c}.
\label{eq:SDE_amp}
\end{equation}
Equivalently, the stochastic Bloch vector $\tilde{\bm r}=(\tilde r_x,\tilde r_y,\tilde r_z)^T$ evolves as
\begin{equation}
\frac{d\tilde{\bm r}}{dt}=\left[\Omega\bm e_x+\beta(t)\bm e_z\right]\times \tilde{\bm r}.
\label{eq:Bloch_lab}
\end{equation}
For an initial ground state with Bloch vector $\bm r_0=(0,0,-1)^T$, the measured excited-state or shelved-state probability is
\begin{equation}
P_e(t)=\frac{1+\langle \tilde r_z(t)\rangle}{2},
\label{eq:prob}
\end{equation}
where the average is over stochastic realizations.

To isolate the noise response, move to the toggling frame of the ideal Rabi drive.  With
\begin{equation}
U_0(t)=\exp[-i\Omega t\sigma_x/2],
\end{equation}
the detuning operator becomes
\begin{equation}
U_0^\dagger(t)\frac{\sigma_z}{2}U_0(t)=\frac{1}{2}\left[\sigma_z\cos\Omega t+\sigma_y\sin\Omega t\right].
\label{eq:toggling_operator}
\end{equation}
In Bloch-vector form,
\begin{equation}
\frac{d\tilde{\bm r}_I}{dt}=\beta(t) A(t)\tilde{\bm r}_I,
\qquad A(t)\bm v=\bm n(t)\times \bm v,
\label{eq:Bloch_I}
\end{equation}
with
\begin{equation}
\bm n(t)=(0,\sin\Omega t,\cos\Omega t)^T.
\label{eq:n_axis}
\end{equation}
The sign in Eq.~\eqref{eq:n_axis} follows from the convention in Eq.~\eqref{eq:toggling_operator}; changing the convention for $U_0$ changes intermediate signs but not laboratory-frame observables.  Equation~\eqref{eq:Bloch_I} shows why driven Rabi decay is not identical to Ramsey dephasing: in the driven frame, detuning noise rotates around a time-dependent axis at the Rabi frequency.

\section{Full finite-time dynamical map}

\subsection{PSD convention and cumulant equation}

We assume that $\beta(t)$ is a zero-mean, stationary, Gaussian process with autocorrelation
\begin{equation}
C(\tau)=\langle \beta(t+\tau)\beta(t)\rangle.
\end{equation}
The two-sided angular-frequency PSD convention is
\begin{equation}
C(\tau)=\int_{-\infty}^{\infty}\frac{d\omega}{2\pi}S_\beta(\omega)e^{i\omega\tau},
\qquad S_\beta(-\omega)=S_\beta(\omega).
\label{eq:PSDconvention}
\end{equation}
To second order in the noise strength, the time-convolution-less cumulant expansion gives
\begin{equation}
\frac{d}{dt}\langle\bm r_I(t)\rangle=K(t)\langle\bm r_I(t)\rangle,
\label{eq:TCL}
\end{equation}
where
\begin{equation}
K(t)=\int_0^t d\tau\, C(\tau)A(t)A(t-\tau).
\label{eq:Ktime}
\end{equation}
The term non-Markovian is used here in a restricted operational sense: the averaged equation is time local, but its generator depends on elapsed time through the finite correlation-memory integral. We do not use this terminology as a claim about CP-divisibility, which would require a separate analysis of the resulting map.


The tensor structure follows from
\begin{equation}
[\bm a]_{\times}[\bm b]_{\times}
=
\bm b\bm a^T-(\bm a\cdot\bm b)I_3 ,
\end{equation}
which gives
\begin{equation}
K(t)=\int_0^t d\tau\, C(\tau)
\left[
\bm n(t-\tau)\bm n^T(t)-\cos(\Omega\tau)I_3
\right].
\label{eq:Kexplicit}
\end{equation}
The diagonal damping terms therefore have the correct negative sign for components perpendicular to the instantaneous noise axis.

Using Eq.~\eqref{eq:PSDconvention}, the same generator is a matrix-valued spectral response,
\begin{equation}
K(t)=\int_{-\infty}^{\infty}\frac{d\omega}{2\pi}S_\beta(\omega)\mathcal{F}(\omega,\Omega,t),
\label{eq:matrixfilter}
\end{equation}
with
\begin{equation}
\mathcal{F}(\omega,\Omega,t)=\int_0^t d\tau\,e^{i\omega\tau}A(t)A(t-\tau).
\end{equation}
This is a matrix filter rather than a single scalar window.  Different Bloch components therefore experience different filtered combinations of the same PSD.

\subsection{Pauli-transfer matrix and average gate error}

Because Eq.~(\ref{eq:TCL}) is linear in the averaged Bloch vector, its solution can be written as a linear map acting on the initial Bloch vector,
\begin{equation}
    \langle \mathbf r_I(t) \rangle = M_I(t)\mathbf r_I(0),
    \label{eq:MI_definition}
\end{equation}
where $M_I(t)$ is the toggling-frame Pauli-transfer matrix of the noise-induced error map. Substituting Eq.~\eqref{eq:MI_definition} into the averaged equation of motion gives
\begin{equation}
    \dot{M}_I(t)=K(t)M_I(t),\qquad M_I(0)=\mathbb{I}_3 .
    \label{eq:MI_eom}
\end{equation}


The laboratory-frame averaged Bloch map is
\begin{equation}
\langle\bm r(t)\rangle=R_x(\Omega t)M_I(t)\bm r(0).
\label{eq:labmap}
\end{equation}
Thus $M_I(t)$ is the Pauli-transfer matrix of the error map relative to the ideal Rabi rotation.  Since the noise is classical detuning noise and the model is unital,
\begin{equation}
\rho(t)=\frac{1}{2}\left[\mathbb{I}+R_x(\Omega t)M_I(t)\bm r(0)\cdot\bm\sigma\right].
\end{equation}
For a unital single-qubit error map, the average gate infidelity relative to the ideal Rabi rotation is
\begin{equation}
\epsilon_{\rm avg}(t)=1-F_{\rm avg}(t)=\frac{3-{\rm Tr}[M_I(t)]}{6}.
\label{eq:gateerror}
\end{equation}
This attributes the same finite-time response used to interpret Rabi decay to a gate-level metric.

\subsection{Scalar secular envelope}

For the specific observable $P_e(t)$, a cycle-averaged secular reduction of the matrix response gives the scalar contrast functional
\begin{equation}
\chi_R(t)=\frac{1}{2}\int_0^t d\tau\,(t-\tau)C(\tau)\cos(\Omega\tau).
\label{eq:chi_corr}
\end{equation}
With Eq.~\eqref{eq:PSDconvention},
\begin{equation}
\chi_R(t)=\int_{-\infty}^{\infty}\frac{d\omega}{2\pi}S_\beta(\omega)F_R(\omega,\Omega,t),
\label{eq:chi_psd}
\end{equation}
where
\begin{equation}
F_R(\omega,\Omega,t)=\frac{1}{8}\left|\int_0^t ds\,e^{i(\omega-\Omega)s}\right|^2+
\frac{1}{8}\left|\int_0^t ds\,e^{i(\omega+\Omega)s}\right|^2.
\label{eq:scalarfilter}
\end{equation}
Equivalently,
\begin{equation}
F_R=\frac{1}{2}\frac{\sin^2[(\omega-\Omega)t/2]}{(\omega-\Omega)^2}
+\frac{1}{2}\frac{\sin^2[(\omega+\Omega)t/2]}{(\omega+\Omega)^2},
\label{eq:scalarfilter_sinc}
\end{equation}
with the appropriate limiting value at the removable singularities.  The scalar Rabi signal is approximated as
\begin{equation}
P_e(t)\simeq \frac{1}{2}\left[1-e^{-\chi_R(t)}\cos(\Omega t+\Delta_R(t))\right],
\label{eq:scalar_signal}
\end{equation}
where $\Delta_R(t)$ represents coherent shifts generated by nonsecular components of the full matrix map.  In the long-time limit,
\begin{equation}
\frac{F_R(\omega,\Omega,t)}{t}\rightarrow \frac{\pi}{4}\left[\delta(\omega-\Omega)+\delta(\omega+\Omega)\right],
\end{equation}
and therefore, for an even two-sided PSD,
\begin{equation}
\frac{\chi_R(t)}{t}\rightarrow \frac{S_\beta(\Omega)}{4}.
\label{eq:longtime_rate}
\end{equation}
The numerical factor is fixed by Eqs.~\eqref{eq:H} and \eqref{eq:PSDconvention}. 

\begin{figure*}[ht!]
    \centering
    \begin{subfigure}[b]{0.48\linewidth}
        \includegraphics[width=\linewidth]{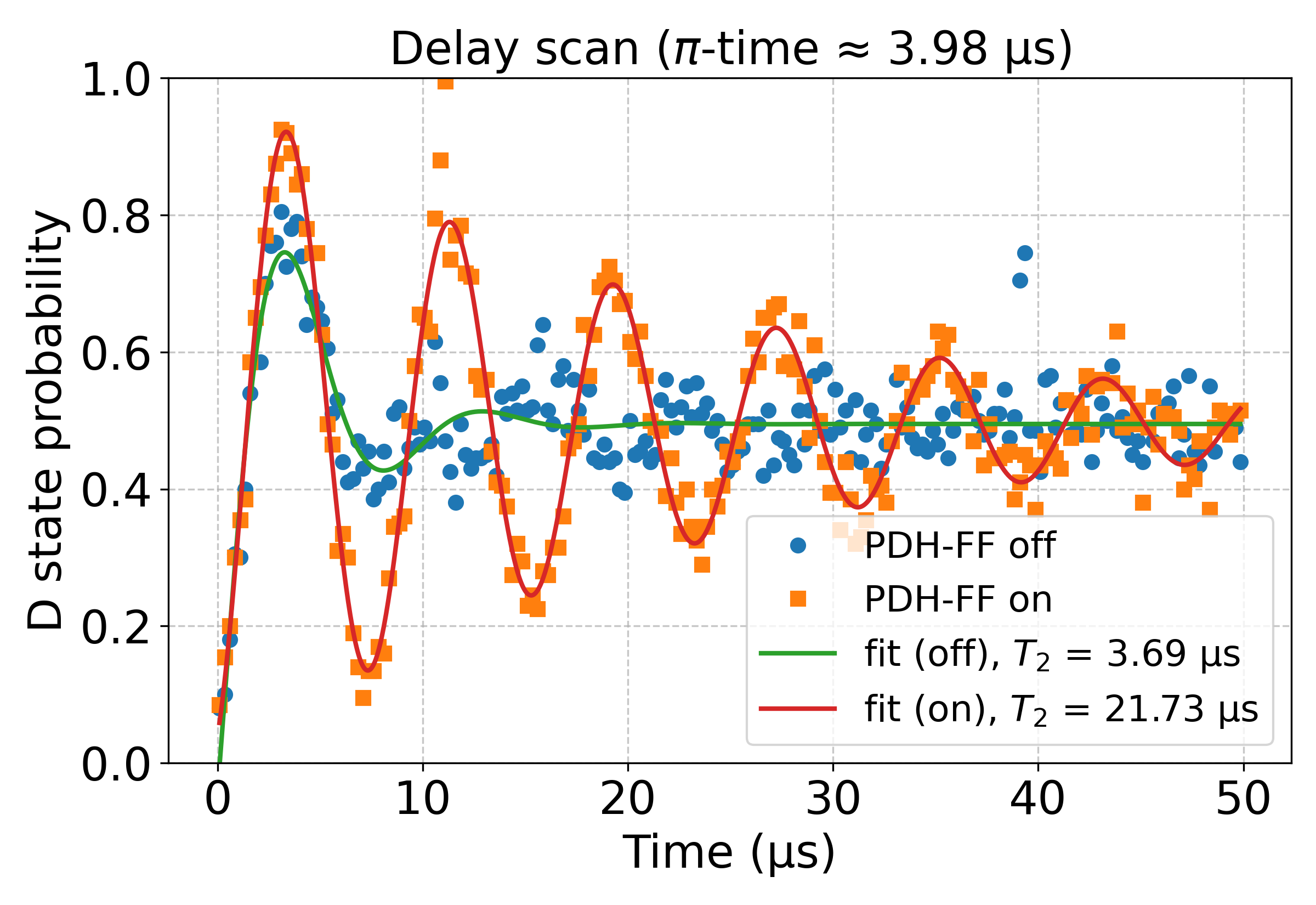}
        \caption{time scan of Rabi oscillation}
        \label{fig:delay_scan_pdhff}
    \end{subfigure}
    \hfill
    \begin{subfigure}[b]{0.48\linewidth}
        \includegraphics[width=\linewidth]{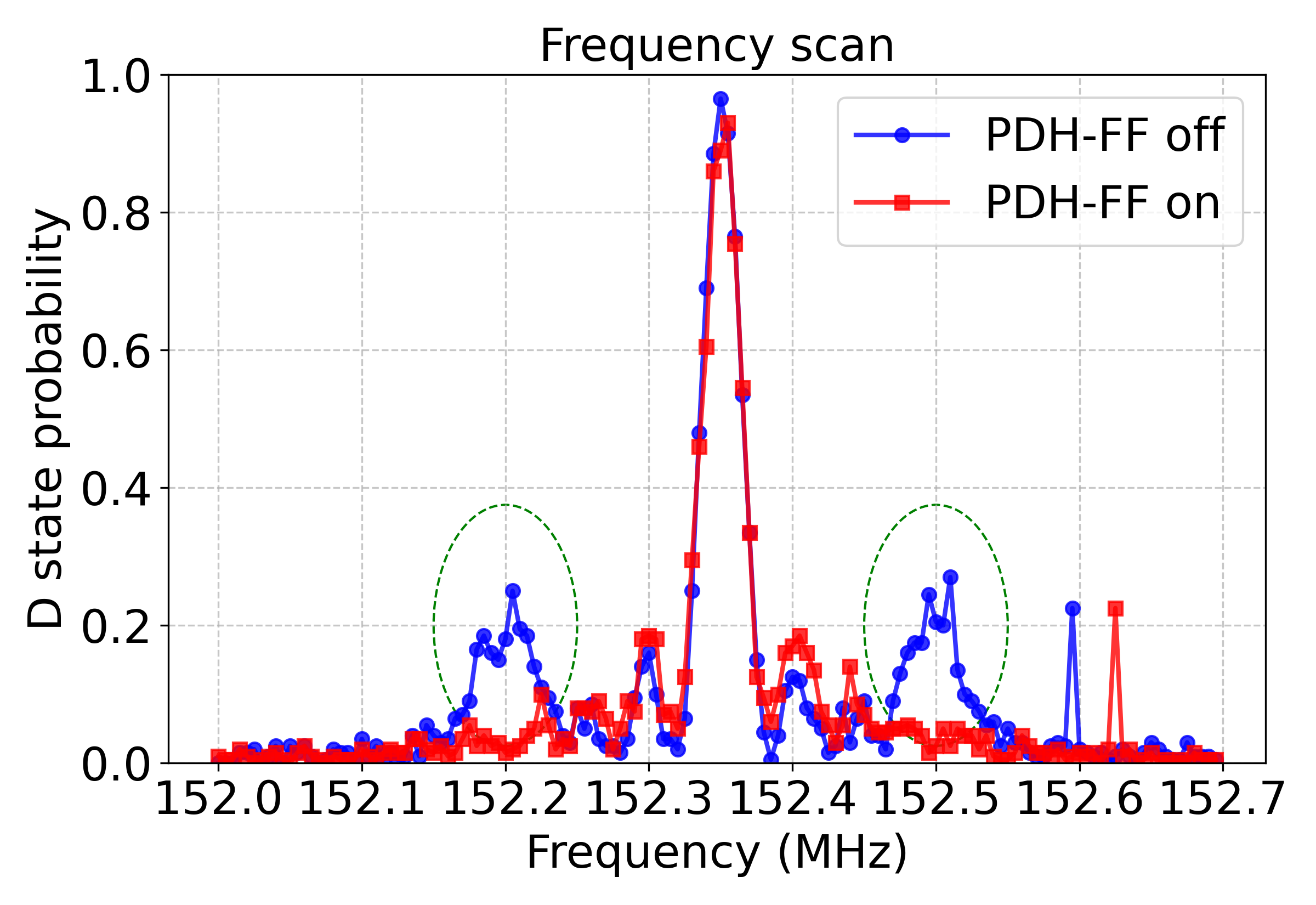}
        \caption{frequency scan of Rabi oscillation}
        \label{fig:freq_scan_pdhff}
    \end{subfigure}

    \caption[Rabi Oscillation measurements]{{\bfseries Rabi Oscillation measurements.} \figureautorefname{~\ref{fig:delay_scan_pdhff}} compare the Rabi oscillations with and without phase noise stabilization and perform $T_2$ fitting. From the data, we observe that $T_2$ is longer when the beam is stabilized. The D-state probability at the first $\pi$-time improves, demonstrating a corresponding enhancement in gate fidelity. \figureautorefname{~\ref{fig:freq_scan_pdhff}} compare the frequency scan excitation spectra with and without phase noise stabilization (pi time used: 22.1us). The additional off-resonant excitation is clearly observable (circled by a dashed green line), with its shape determined by the servo noise and locking bandwidth.}
    \label{fig:T2_fitting}
\end{figure*}

\section{Impact on the ion qubit}

\subsection{Experimental setup}
A single \({}^{138}\mathrm{Ba}^{+}\) ion is confined in a linear Paul trap housed in an ultra-high-vacuum (UHV) chamber.  The ion is Doppler cooled using a $493~$nm laser, red-detuned from the \(\mathrm{S}_{1/2} \rightarrow \mathrm{P}_{1/2}\) transition, together with a $650~$nm repump laser to return population from the meta-stable \(\mathrm{D}_{3/2}\) state. 
A narrow-linewidth (\(\sim 100~\mathrm{Hz}\)) 1762~nm laser, addressing the \(\mathrm{S}_{1/2} \rightarrow \mathrm{D}_{5/2}\) quadrupole transition, is employed to drive Rabi oscillations after the ion is optically pumped to the \(\mathrm{S}_{1/2},\, m_{J} = +\tfrac{1}{2}\) state using a \(\sigma^{+}\)-polarized 493~nm beam. 
The quantization axis is defined by a magnetic field oriented at \(45^{\circ}\) with respect to the propagation direction of the horizontally polarized $1762~$nm beam. The magnetic-field strength is chosen such that the $S_{1/2}$ Zeeman splitting is approximately \(11~\mathrm{MHz}\). Population in $D_{5/2}$ is detected via fluorescence on the $493~$nm dipole transition, collected on a PMT (quantum efficiency$\sim50\%$), enabling fast, high-fidelity discrimination of the shelved $D_{5/2}$ state. Population in $D_{5/2}$ is deshelved with $614~$nm light for subsequent experiments. For further details of the setup, please check \cite{yum2017optical}.

\subsection{Measurements}

The Rabi frequency is first calibrated by scanning the pulse duration and identifying the first maximum (\(\pi\)-pulse). To quantify the phase stability of the laser, we fit the Rabi oscillation using sine damp function as in Eq.~\eqref{eq:rabi_decay} to estimate the coherence time.  

\begin{equation}
P_D(t) = A e^{-t/T_2} \sin^2\left(\frac{\Omega t}{2}+\phi\right) + B ,
\label{eq:rabi_decay}
\end{equation}

where \( T_2 \) characterizes dephasing due to residual phase noise. The coherence time \( T_2 \) extracted from Rabi oscillations is expected to follow \( T_2^{-1} \propto S_{\phi}(f) \) integrated over the qubit’s filter function, which weights phase noise near the Rabi frequency. Thus, suppression of the servo-bump noise directly improves \( T_2 \) for resonant Rabi drives in this frequency range.

Also, we test the stabilized beam by scanning the frequency of the driving laser to see the effect on the off-resonant excitation. The excitation probability was recorded while detuning the $1762~$nm laser over \(\pm 700~\mathrm{kHz}\) in \(5~\mathrm{kHz}\) steps. 
The off-resonant excitation shoulder corresponds to the servo-bump-induced phase 
modulation near \(132~\mathrm{kHz}\), consistent with the measured noise spectrum.
 
\autoref{fig:T2_fitting} shows one example of our results when the Rabi frequency is close to the servo-bump frequency. Each data point represents the average of $200$ experiments, and the sequence is repeated for each delay to extract the full oscillation. 
The fitted coherence time \(T_2\) improves from \(3.69 \pm 0.35~\mu\mathrm{s}\) (un-stabilized) to \(21.73 \pm 0.87~\mu\mathrm{s}\) (stabilized), corresponding to an enhancement factor of approximately 6 times in this example.
By reducing the servo-bump phase noise, the D-state probability at the first
\(\pi\)-time increases from \(0.641 \pm 0.014\) to \(0.870 \pm 0.009\),
demonstrating improved \(\pi\)-pulse state-transfer performance. This
single-state measurement is distinct from a full average gate fidelity, which
requires averaging over all input states. To estimate the corresponding phase-noise-limited average gate fidelity, we use
the finite-time Pauli-transfer matrix introduced in Sec.~IV. For a resonant
\(\pi\)-pulse with \(t_\pi=\pi/\Omega\), the average gate fidelity limited by
laser phase noise is
\[
F_{\mathrm{avg}}^{(\phi)}(t_\pi)
=
\frac{3+\mathrm{Tr}[M_I(t_\pi)]}{6}.
\]
Using the DSHI-measured phase-noise spectra with PDH feedforward disabled and
enabled, together with the measured Rabi decay to calibrate the overall noise
scale, we estimate that, for operation near the servo-bump frequency, PDH
feedforward improves the phase-noise-limited average \(\pi\)-gate fidelity by
\[
\Delta F_{\mathrm{avg}}^{(\phi)} \simeq 0.22.
\]
A broader Rabi-frequency scan is presented in Appendix~\ref{app:fidelity_frequency_scan}, where the same calibrated finite-time model is used to evaluate the phase-noise-limited $\pi$-gate infidelity from (20) to (450~$\mathrm{kHz}$) and quantify the feedforward improvement around the servo-bump region. Since \(X_\pi\) and \(Y_\pi\) rotations differ only by the optical drive phase,
and the phase noise enters as a stochastic detuning term proportional to
\(\sigma_z\), the same estimate applies to both \(X_\pi\) and \(Y_\pi\) gates
within this model.

\begin{figure}[ht!]
    \centering
    \includegraphics[width=1.0\linewidth]{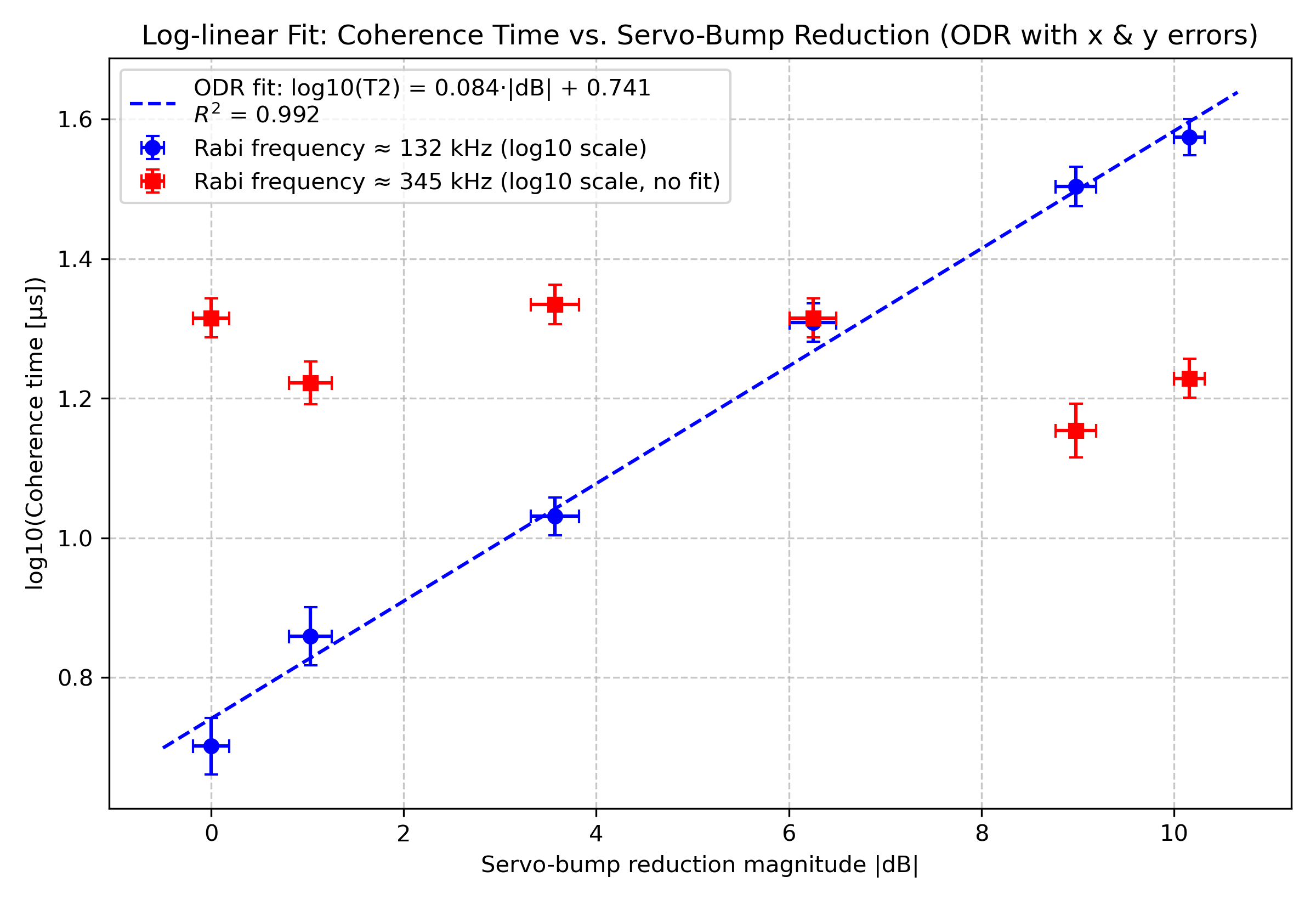}
    \caption[$T_2$ vs servo-bump noise reduction]{
    {\bfseries Log-linear scaling of $T_2$ coherence time with servo-bump phase-noise reduction.} 
    The plot shows the measured coherence time $T_2$ as a function of the servo-bump phase-noise reduction (in dB). 
    A log-linear fit shows an improvement factor of approximately $1.21\times$ in $T_2$ per dB of noise suppression. 
    Blue points (Rabi frequency $\approx 132~\mathrm{kHz}$) are used for the fit, while red points ($\approx 345~\mathrm{kHz}$) are shown for comparison.
    }
    \label{fig:T2_dB}
\end{figure}

We also summarize the relationship between the coherence time and the servo-bump phase-noise reduction in \figureautorefname{ \ref{fig:T2_dB}}. 
The data shows a log-linear dependence, fitted by
\[
\log_{10}(T_2) = 0.084 \times |\Delta_{\mathrm{dB}}| + 0.741,
\]
with a coefficient of determination $R^2 = 0.992$. This corresponds to an improvement factor of approximately 1.21 times in $T_2$ per dB of servo-bump noise suppression. 
The blue data points correspond to a Rabi frequency of $\approx 132~\mathrm{kHz}$ and are included in the fit, 
while the red points correspond to $\approx 345~\mathrm{kHz}$ and are shown for comparison. 
As expected, the effect is minimal when the Rabi frequency does not overlap with the servo-bump phase-noise frequency, which is experimentally confirmed in our results.

\section{\label{sec:discussion}Discussion}

We experimentally demonstrated that phase noise stabilization based on the PDH feedforward technique can reduce $15~$dB servo-bump phase noise near the bump peak frequency created by the locking setup while maintaining the linewidth of the seed laser. Importantly, our results confirm that this method can be successfully applied to a laser driving a quadrupole transition in an optical qubit gate —specifically, the $1762~$nm $\mathrm{S}_{1/2} \leftrightarrow \mathrm{D}_{5/2}$ transition in ${}^{138}\mathrm{Ba}^{+}$. We verified that the suppression directly translates into a measurable improvement in qubit coherence time factor of approximately $1.21\times$ in $T_2$ per dB of servo-bump noise suppression. This provides the first experimental verification that PDH feedforward method can be integrated into optical qubit operations and used to enhance trapped-ion gate performance.

Several methods have been developed to reduce the servo-bump phase noise arising from feedback bandwidth limitations in PDH locking~\cite{chao2024pound,chao2025robust,krinner2024low,hald2005efficient,akerman2015universal,levine2018high,nazarova2008low,labaziewicz2007compact,fluhmann2019encoding}. Among them, the PDH feedforward method offers a direct and power-efficient approach, utilizing the residual PDH error signal—encoding instantaneous phase fluctuations from the locking system—and applying it as a compensating modulation to a high-speed actuator~\cite{chao2024pound,chao2025robust}. By properly tuning the gain and matching the delay, the servo-bump noise beyond the feedback bandwidth can be strongly suppressed. In our case, optimal cancellation was achieved around a relatively low bump frequency of 132 kHz, demonstrating the method’s flexibility compared with prior implementations operating at MHz frequencies.

The limitation of the transmitted power in this method arises from the fiber EOM, which can handle only a few hundred milliwatts in the infrared region (e.g., PM1750). For visible wavelengths, the allowable power decreases to only a few milliwatts, necessitating the use of a broadband crystal EOM instead with the requirement of a high-voltage amplifier.

An alternative approach uses the beat signal between the cavity-filtered light and the cavity-stabilized laser to reduce servo bumps via a high-speed electro-optical feedforward loop~\cite{li2022active}. In another study~\cite{krinner2024low}, the authors present a self-injection locked diode laser system that leverages a high-finesse cavity. This cavity serves two key functions: providing a stable resonance frequency and acting as a low-pass filter for phase noise beyond the cavity linewidth. As a result, the system achieves low phase noise from DC up to the injection lock limit. Another solution uses cavity-filtered light~\cite{hald2005efficient,akerman2015universal,levine2018high,nazarova2008low,labaziewicz2007compact,fluhmann2019encoding}. However, the power transmitted through a narrow-linewidth cavity is usually restricted to a few milliwatts or less to avoid coating damage due to the buildup of intense optical power inside the cavity. 

We chose to use the PDH feedforward method because the transmitted power through our ultra-stable cavity is only a few microwatts, making cavity filtering impractical. The PDH feedforward approach thus provides a compact, power-preserving solution that does not sacrifice optical throughput. Furthermore, our trapped-ion experiments confirm that the noise suppression achieved by this method leads to a direct improvement in coherence time, validating its relevance for optical qubit gate operations. As for example \cite{akerman2015universal}, they have observed that the fidelity of the Mølmer–Sørensen (MS) entangling gate is severely compromised by the presence of servo-bumps near the motional frequency. It will be more severe when considering long chain approach to scaling where frequency crowding is a challenge. Therefore, building on the observed improvement in single-qubit coherence, future work will focus on assessing the influence of servo-bump phase-noise suppression on two-qubit gate fidelity. This will provide a more complete understanding of its role in achieving high-fidelity quantum logic operations in trapped-ion processors. 

\section{\label{sec:conclusion
}Conclusion}

In this work, we developed a $1762~$nm laser system incorporating phase-noise stabilization via the PDH feedforward method, which suppresses the $15~$dB servo-bump phase noise arising from the locking system. Using delayed self-heterodyne interferometry (DSHI), we analyzed the phase-noise spectrum and linewidth of the stabilized beam and found that the feedforward method does not significantly broaden the laser linewidth while significantly supressing the servo-bump. We further tested the stabilized laser in our trapped-ion system, observing clear improvements in the coherence time and gate fidelity of a single-ion optical qubit. These results demonstrate the effectiveness of the PDH feedforward approach and provide the first experimental verification that it can be successfully applied to lasers used for optical qubit gate operations.

\section*{Data Availability}

The data and analysis code supporting the figures in this manuscript are
available from the corresponding author upon reasonable request.

\section*{Acknowledgment}
This work is supported under the National Research Foundation (NRF), Singapore grant within National Quantum Processor Initiative (grant no. S24Q4D0001) and CQT core grant (A-8003276-00-00) administered by National Quantum Office, A*star. The authors thank Clarence Liu, Jasper Phua, Subhadeep Chakraborty, and Low Pei Jiang, along with the other members of the trapped-ion laboratory at the Centre for Quantum Technologies, for useful discussions and
experimental support.

\appendix

\section{Derivation and validity of the second-order response generator}
\label{app:generator_derivation}

This appendix gives the derivation of the finite-time response generator used in the main text.  In the toggling frame of the ideal Rabi drive, the trajectory-level Bloch vector obeys
\begin{equation}
\frac{d\tilde{\bm r}_I}{dt}
=
\beta(t)A(t)\tilde{\bm r}_I ,
\label{eq:app_bloch_I}
\end{equation}
where \(A(t)\bm v=\bm n(t)\times \bm v\), with
\begin{equation}
\bm n(t)=
\begin{pmatrix}
0\\
\sin\Omega t\\
\cos\Omega t
\end{pmatrix}.
\end{equation}
Equation~\eqref{eq:app_bloch_I} has the formal solution
\begin{equation}
\tilde{\bm r}_I(t)
=
\mathcal{T}
\exp\left[
\int_0^t ds\,\beta(s)A(s)
\right]\bm r(0),
\label{eq:app_formal_solution}
\end{equation}
where \(\mathcal{T}\) denotes time ordering.

Expanding Eq.~\eqref{eq:app_formal_solution} to second order gives
\begin{align}
\tilde{\bm r}_I(t)
=&
\left[
I_3
+
\int_0^t ds_1\,\beta(s_1)A(s_1)
\right.
\nonumber\\
&\left.
+
\int_0^t ds_1
\int_0^{s_1} ds_2\,
\beta(s_1)\beta(s_2)A(s_1)A(s_2)
+\cdots
\right]\bm r(0).
\label{eq:app_dyson}
\end{align}
For zero-mean noise, the first-order contribution vanishes after averaging.  The leading nonzero contribution is controlled by the two-point correlation function
\begin{equation}
C(s_1-s_2)=\langle \beta(s_1)\beta(s_2)\rangle .
\end{equation}
The corresponding second-order time-convolutionless equation is
\begin{equation}
\frac{d}{dt}\langle \bm r_I(t)\rangle
=
K(t)\langle \bm r_I(t)\rangle
+
O(\beta^4),
\label{eq:app_tcl}
\end{equation}
with
\begin{equation}
K(t)
=
\int_0^t ds\,
\langle \beta(t)\beta(s)\rangle A(t)A(s).
\label{eq:app_K_s}
\end{equation}
For stationary noise, \(\langle \beta(t)\beta(s)\rangle=C(t-s)\).  Setting \(\tau=t-s\) gives
\begin{equation}
K(t)
=
\int_0^t d\tau\,C(\tau)A(t)A(t-\tau),
\label{eq:app_K_tau}
\end{equation}
which is Eq.~\eqref{eq:Ktime} of the main text.

The explicit matrix form follows from the cross-product identity
\begin{equation}
[\bm a]_{\times}[\bm b]_{\times}
=
\bm b\bm a^T-(\bm a\cdot\bm b)I_3 .
\label{eq:app_cross_identity}
\end{equation}
Using \(A(t)=[\bm n(t)]_{\times}\),
\begin{equation}
A(t)A(t-\tau)
=
\bm n(t-\tau)\bm n^T(t)
-
[\bm n(t)\cdot\bm n(t-\tau)]I_3 .
\end{equation}
Since
\begin{align}
\bm n(t)\cdot\bm n(t-\tau)
&=
\sin\Omega t\sin\Omega(t-\tau)
+
\cos\Omega t\cos\Omega(t-\tau)
\nonumber\\
&=
\cos(\Omega\tau),
\end{align}
the generator can be written as
\begin{equation}
K(t)
=
\int_0^t d\tau\,C(\tau)
\left[
\bm n(t-\tau)\bm n^T(t)
-
\cos(\Omega\tau)I_3
\right].
\label{eq:app_K_explicit}
\end{equation}
At \(\tau=0\), the matrix inside the brackets is \(\bm n\bm n^T-I_3\), which has eigenvalue zero along \(\bm n\) and eigenvalue \(-1\) in the two directions transverse to \(\bm n\).  This explains the negative sign of the damping terms.

The approximation in Eq.~\eqref{eq:app_tcl} is perturbative.  A useful scale for its validity is the accumulated second cumulant
\begin{equation}
\Lambda(t)
=
\int_0^t ds
\int_0^s ds'\,
|C(s-s')|,
\label{eq:app_validity}
\end{equation}
up to factors set by the norm of the rotation generators.  The second-order response map is expected to be reliable when this quantity is small to moderate and higher cumulants are negligible.  For stronger noise or longer evolution times, the truncated map may lose complete positivity.  In that regime, direct Monte-Carlo propagation of the stochastic Bloch equation should be used as the positivity-preserving reference.

\section{Scalar Rabi filter and long-time limit}
\label{app:scalar_filter}

This appendix derives the scalar Rabi-envelope filter used in the main text and fixes the numerical factor in the long-time limit.  The full response map is matrix valued, but for the measured Rabi contrast one may obtain a scalar envelope by applying a secular approximation to the driven-frame noise response.

For a stationary Gaussian detuning process, define
\begin{equation}
C(\tau)=\langle \beta(t+\tau)\beta(t)\rangle ,
\end{equation}
with the two-sided angular-frequency PSD convention
\begin{equation}
C(\tau)
=
\int_{-\infty}^{\infty}
\frac{d\omega}{2\pi}
S_\beta(\omega)e^{i\omega\tau},
\qquad
S_\beta(-\omega)=S_\beta(\omega).
\end{equation}
In the toggling frame of the Rabi drive, the detuning noise rotates at the Rabi frequency.  Keeping the secular contribution to the Rabi contrast gives the scalar decay exponent
\begin{equation}
\chi_R(t)
=
\frac{1}{2}
\int_0^t d\tau\,
(t-\tau)C(\tau)\cos(\Omega\tau).
\label{eq:app_chi_scalar}
\end{equation}
This is the finite-time Rabi-window functional in the time domain.

Substituting the PSD representation of \(C(\tau)\), Eq.~\eqref{eq:app_chi_scalar} can be written as
\begin{equation}
\chi_R(t)
=
\int_{-\infty}^{\infty}
\frac{d\omega}{2\pi}
S_\beta(\omega)F_R(\omega,\Omega,t),
\label{eq:app_chi_freq}
\end{equation}
where the scalar filter is
\begin{equation}
F_R(\omega,\Omega,t)
=
\frac{1}{2}
\int_0^t d\tau\,
(t-\tau)\cos(\omega\tau)\cos(\Omega\tau).
\label{eq:app_FR_integral}
\end{equation}
Using
\begin{equation}
\cos(\omega\tau)\cos(\Omega\tau)
=
\frac{1}{2}
\left[
\cos((\omega-\Omega)\tau)
+
\cos((\omega+\Omega)\tau)
\right],
\end{equation}
and the identity
\begin{equation}
\left|
\int_0^t ds\,e^{ias}
\right|^2
=
2\int_0^t d\tau\,(t-\tau)\cos(a\tau),
\label{eq:app_window_identity}
\end{equation}
one obtains
\begin{equation}
F_R(\omega,\Omega,t)
=
\frac{1}{8}
\left|
\int_0^t ds\,e^{i(\omega-\Omega)s}
\right|^2
+
\frac{1}{8}
\left|
\int_0^t ds\,e^{i(\omega+\Omega)s}
\right|^2 .
\label{eq:app_FR_exp}
\end{equation}
Evaluating the finite-time integral,
\begin{equation}
\int_0^t ds\,e^{ias}
=
e^{iat/2}
\frac{2\sin(at/2)}{a},
\end{equation}
gives
\begin{equation}
F_R(\omega,\Omega,t)
=
\frac{1}{2}
\frac{\sin^2[(\omega-\Omega)t/2]}{(\omega-\Omega)^2}
+
\frac{1}{2}
\frac{\sin^2[(\omega+\Omega)t/2]}{(\omega+\Omega)^2}.
\label{eq:app_FR_sinc}
\end{equation}
The values at \(\omega=\pm\Omega\) are understood by taking the removable limits.

The long-time limit follows from the distribution identity
\begin{equation}
\lim_{t\to\infty}
\frac{1}{t}
\left|
\int_0^t ds\,e^{ias}
\right|^2
=
2\pi\delta(a),
\label{eq:app_delta_identity}
\end{equation}
or equivalently
\begin{equation}
\lim_{t\to\infty}
\frac{\sin^2(at/2)}{t a^2}
=
\frac{\pi}{2}\delta(a).
\end{equation}
Therefore,
\begin{equation}
\frac{F_R(\omega,\Omega,t)}{t}
\rightarrow
\frac{\pi}{4}
\left[
\delta(\omega-\Omega)
+
\delta(\omega+\Omega)
\right].
\label{eq:app_FR_long}
\end{equation}
Substituting Eq.~\eqref{eq:app_FR_long} into Eq.~\eqref{eq:app_chi_freq} gives
\begin{align}
\frac{\chi_R(t)}{t}
&\rightarrow
\frac{\pi}{4}
\int_{-\infty}^{\infty}
\frac{d\omega}{2\pi}
S_\beta(\omega)
\left[
\delta(\omega-\Omega)
+
\delta(\omega+\Omega)
\right]
\nonumber\\
&=
\frac{1}{8}
\left[
S_\beta(\Omega)+S_\beta(-\Omega)
\right].
\end{align}
For an even two-sided PSD,
\begin{equation}
\frac{\chi_R(t)}{t}
\rightarrow
\frac{S_\beta(\Omega)}{4}.
\label{eq:app_T2_limit}
\end{equation}
Thus, if the Rabi contrast envelope is written as \(\exp[-t/T_2]\), the long-time narrow-filter limit gives
\begin{equation}
T_2^{-1}
=
\frac{S_\beta(\Omega)}{4}.
\end{equation}
This relation is not a finite-time identity.  At finite duration, the Rabi experiment samples the PSD through the full window \(F_R(\omega,\Omega,t)\), whose width is of order \(1/t\).

\section{Average gate fidelity from the toggling-frame Pauli-transfer matrix} \label{app:avg_gate_fidelity} This appendix derives the relation between the finite-time Pauli-transfer matrix $M_I(t)$ and the average gate infidelity used in the main text. In the toggling frame of the ideal Rabi drive, the averaged Bloch vector evolves as \begin{equation} \langle \mathbf r_I(t) \rangle = M_I(t)\mathbf r_I(0), \label{eq:app_MI_map} \end{equation} where $M_I(t)$ is the noise-induced error map relative to the ideal resonant Rabi rotation. Returning to the laboratory rotating frame gives \begin{equation} \langle \mathbf r(t) \rangle = R_x(\Omega t)M_I(t)\mathbf r(0). \end{equation} Thus $M_I(t)$, rather than the full map $R_x(\Omega t)M_I(t)$, determines the gate error relative to the ideal operation. For a qubit state written in Bloch form, \begin{equation} \rho=\frac{1}{2}\left(I+\mathbf r\cdot\boldsymbol{\sigma}\right), \end{equation} a unital error map transforms the Bloch vector according to \begin{equation} \mathbf r \longmapsto M_I(t)\mathbf r . \end{equation} The phase-noise model considered here is unital because classical detuning noise averages over rotations and does not displace the center of the Bloch sphere. Consider an arbitrary pure input state with $|\mathbf r|=1$. In the toggling-frame error picture, the ideal output has Bloch vector $\mathbf r$, whereas the noisy output has Bloch vector $M_I(t)\mathbf r$. The fidelity between a pure qubit state with Bloch vector $\mathbf r$ and a general qubit state with Bloch vector $\mathbf r'$ is \begin{equation} F(\mathbf r,\mathbf r') = \frac{1+\mathbf r\cdot\mathbf r'}{2}. \end{equation} Therefore, for the error map $M_I(t)$, \begin{equation} F(\mathbf r) = \frac{1+\mathbf r^T M_I(t)\mathbf r}{2}. \end{equation} The average gate fidelity is obtained by averaging this expression uniformly over all pure input states on the Bloch sphere: \begin{equation} F_{\mathrm{avg}}(t) = \left\langle \frac{1+\mathbf r^T M_I(t)\mathbf r}{2} \right\rangle_{\mathbf r}. \end{equation} Using the spherical average \begin{equation} \left\langle r_i r_j \right\rangle_{\mathbf r} = \frac{1}{3}\delta_{ij}, \end{equation} we obtain \begin{align} \left\langle \mathbf r^T M_I(t)\mathbf r \right\rangle_{\mathbf r} &= \sum_{i,j} \left[M_I(t)\right]_{ij} \left\langle r_i r_j \right\rangle_{\mathbf r} \\ &= \frac{1}{3}\sum_i \left[M_I(t)\right]_{ii} \\ &= \frac{1}{3}\mathrm{Tr}\left[M_I(t)\right]. \end{align} Thus \begin{equation} F_{\mathrm{avg}}(t) = \frac{1}{2} + \frac{1}{2} \frac{\mathrm{Tr}\left[M_I(t)\right]}{3} = \frac{3+\mathrm{Tr}\left[M_I(t)\right]}{6}. \end{equation} The corresponding average gate infidelity is \begin{equation} \epsilon_{\mathrm{avg}}(t) = 1-F_{\mathrm{avg}}(t) = \frac{3-\mathrm{Tr}\left[M_I(t)\right]}{6}. \label{eq:app_avg_gate_error} \end{equation} For a $\pi$ pulse, the gate time is $t_g=\pi/\Omega$, and the phase-noise-limited average gate infidelity is therefore \begin{equation} \epsilon_{\mathrm{avg}}^{(\phi)} = \frac{3-\mathrm{Tr}\left[M_I(t_g)\right]}{6}. \end{equation} This quantity is distinct from the state-transfer error $1-P_e(t_g)$ measured for a single initial state. The latter probes one specific input state, while Eq.~\eqref{eq:app_avg_gate_error} averages the error over all pure input states on the Bloch sphere.

In the numerical estimate used in the main text, the DSHI spectra are converted
from RF offset frequency \(f=|f_{\mathrm{RF}}-f_{\mathrm{EOM}}|\), where
\(f_{\mathrm{EOM}}=7~\mathrm{MHz}\). The short-delay DSHI response is corrected
using
\[
H_\tau(f)=4\sin^2(\pi f\tau),
\]
with \(\tau\simeq75~\mathrm{ns}\) for the \(15~\mathrm{m}\) fiber. The resulting
phase-noise spectrum is converted to detuning-noise PSD using
\[
S_\beta(2\pi f)=(2\pi f)^2 S_\phi(f).
\]
The finite-time map \(M_I(t_\pi)\) is then obtained by integrating
Eq.~\eqref{eq:MI_eom}.

\section{Derivation of the scalar Rabi-envelope functional}
\label{app:scalar_rabi_envelope}

This appendix derives the scalar contrast functional used to describe the
finite-time decay of resonant Rabi oscillations under detuning noise.  In the
toggling frame of the ideal Rabi drive, the stochastic Hamiltonian is
\begin{equation}
    H_I(t)
    =
    \frac{\hbar}{2}\beta(t)
    \left[
        \sigma_z\cos(\Omega t)
        +
        \sigma_y\sin(\Omega t)
    \right],
    \label{eq:app_HI_rabi_noise}
\end{equation}
where $\beta(t)=\dot{\phi}(t)$ is the detuning noise generated by laser phase
fluctuations.  Equivalently, the instantaneous noise axis in the toggling frame
is
\begin{equation}
    \mathbf n(t)
    =
    \begin{pmatrix}
        0\\
        \sin(\Omega t)\\
        \cos(\Omega t)
    \end{pmatrix}.
    \label{eq:app_noise_axis}
\end{equation}
Thus detuning noise is modulated at the Rabi frequency $\Omega$ in the driven
frame.

For weak zero-mean Gaussian noise, the slowly varying Rabi contrast can be
written as
\begin{equation}
    \mathcal C_R(t)=e^{-\chi_R(t)},
    \label{eq:app_contrast_envelope}
\end{equation}
where $\chi_R(t)$ is obtained from the second cumulant of the driven-frame
noise.  In the scalar secular approximation, the relevant stochastic
phase-like quadrature is
\begin{equation}
    \theta_R(t)
    =
    \int_0^t dt_1\,\beta(t_1)\cos(\Omega t_1).
    \label{eq:app_theta_R}
\end{equation}
This expression captures the component of the detuning noise that is mixed down
by the Rabi drive into the slowly varying contrast envelope.  The remaining
quadrature gives nonsecular oscillatory corrections and small coherent shifts,
which are not included in the scalar envelope.

For a zero-mean Gaussian variable, the cumulant identity gives
\begin{equation}
    \left\langle e^{i\theta_R(t)}\right\rangle
    =
    \exp\!\left[
        -\frac{1}{2}
        \left\langle \theta_R^2(t)\right\rangle
    \right].
    \label{eq:app_gaussian_cumulant}
\end{equation}
Comparing Eqs.~\eqref{eq:app_contrast_envelope} and
\eqref{eq:app_gaussian_cumulant}, the Rabi-envelope exponent is
\begin{equation}
    \chi_R(t)
    =
    \frac{1}{2}
    \left\langle \theta_R^2(t)\right\rangle .
    \label{eq:app_chi_from_variance}
\end{equation}
Using Eq.~\eqref{eq:app_theta_R}, we obtain
\begin{align}
    \left\langle \theta_R^2(t)\right\rangle
    &=
    \int_0^t dt_1
    \int_0^t dt_2\,
    \left\langle
        \beta(t_1)\beta(t_2)
    \right\rangle
    \nonumber\\
    &\qquad\times
    \cos(\Omega t_1)\cos(\Omega t_2).
    \label{eq:app_theta_variance}
\end{align}
For stationary detuning noise, define
\begin{equation}
    C(t_1-t_2)
    =
    \left\langle
        \beta(t_1)\beta(t_2)
    \right\rangle .
    \label{eq:app_detuning_correlation}
\end{equation}
Then
\begin{align}
    \chi_R(t)
    &=
    \frac{1}{2}
    \int_0^t dt_1
    \int_0^t dt_2\,
    C(t_1-t_2)
    \nonumber\\
    &\qquad\times
    \cos(\Omega t_1)\cos(\Omega t_2).
    \label{eq:app_chi_coscos}
\end{align}

Next, use the trigonometric identity
\begin{align}
    \cos(\Omega t_1)\cos(\Omega t_2)
    &=
    \frac{1}{2}
    \cos\!\left[\Omega(t_1-t_2)\right]
    \nonumber\\
    &\quad+
    \frac{1}{2}
    \cos\!\left[\Omega(t_1+t_2)\right].
    \label{eq:app_cos_product_identity}
\end{align}
Substituting Eq.~\eqref{eq:app_cos_product_identity} into
Eq.~\eqref{eq:app_chi_coscos} gives
\begin{align}
    \chi_R(t)
    &=
    \frac{1}{4}
    \int_0^t dt_1
    \int_0^t dt_2\,
    C(t_1-t_2)
    \cos\!\left[\Omega(t_1-t_2)\right]
    \nonumber\\
    &\quad+
    \frac{1}{4}
    \int_0^t dt_1
    \int_0^t dt_2\,
    C(t_1-t_2)
    \cos\!\left[\Omega(t_1+t_2)\right].
    \label{eq:app_chi_secular_nonsecular}
\end{align}
The second term oscillates rapidly with the sum time $t_1+t_2$ and does not
give a secular contribution to the slowly varying Rabi envelope.  Keeping only
the secular term gives
\begin{align}
    \chi_R(t)
    &=
    \frac{1}{4}
    \int_0^t dt_1
    \int_0^t dt_2\,
    C(t_1-t_2)
    \nonumber\\
    &\qquad\times
    \cos\!\left[\Omega(t_1-t_2)\right].
    \label{eq:app_chi_double_integral}
\end{align}
The factor $1/4$ therefore comes from the product of the Gaussian cumulant
factor $1/2$ and the trigonometric factor $1/2$.

We now reduce the double integral to a single finite-time window.  Define
\begin{align}
    \mathcal I(t)
    &\equiv
    \int_0^t dt_1
    \int_0^t dt_2\,
    C(t_1-t_2)
    \nonumber\\
    &\qquad\times
    \cos\!\left[\Omega(t_1-t_2)\right].
    \label{eq:app_I_definition}
\end{align}
The integrand depends only on the time difference
\begin{equation}
    \tau=t_1-t_2 .
\end{equation}
For a fixed positive separation $\tau$, the number of pairs $(t_1,t_2)$ inside
the square region $0<t_1,t_2<t$ is $t-\tau$.  Since $C(\tau)$ is even for
stationary real noise and $\cos(\Omega\tau)$ is also even, the square integral
can be rewritten as
\begin{align}
    \mathcal I(t)
    &=
    2\int_0^t d\tau\,
    (t-\tau)C(\tau)\cos(\Omega\tau).
    \label{eq:app_double_to_single_window}
\end{align}
Substituting Eq.~\eqref{eq:app_double_to_single_window} into
Eq.~\eqref{eq:app_chi_double_integral} gives
\begin{equation}
    \chi_R(t)
    =
    \frac{1}{2}
    \int_0^t d\tau\,
    (t-\tau)C(\tau)\cos(\Omega\tau).
    \label{eq:app_chi_single_integral}
\end{equation}
This is the scalar finite-time Rabi-window functional used in the main text.

The factor $(t-\tau)$ is the finite-time window: noise correlations at short
time separation occur for many pairs of times inside the Rabi pulse, whereas
correlations near $\tau=t$ occur for only a few pairs.  The factor
$\cos(\Omega\tau)$ is the Rabi modulation in the toggling frame.  Therefore,
Eq.~\eqref{eq:app_chi_single_integral} shows that resonant Rabi oscillations
probe detuning noise filtered around the Rabi frequency.

With this scalar envelope, the measured excited-state probability for an
initial ground state is approximated as
\begin{equation}
    P_e(t)
    \simeq
    \frac{1}{2}
    \left[
        1-e^{-\chi_R(t)}
        \cos\!\left(\Omega t+\Delta_R(t)\right)
    \right],
    \label{eq:app_scalar_rabi_signal}
\end{equation}
where $\Delta_R(t)$ denotes small coherent shifts generated by nonsecular
terms.  This scalar expression is a reduced description of the measured Rabi
contrast; the full gate-level error map is instead obtained from the matrix
response $M_I(t)$.

\section{Phase correction in the scalar Rabi signal}
\label{app:rabi_phase_shift}

This appendix defines the phase correction $\Delta_R(t)$ that appears in the
scalar Rabi signal.  The scalar envelope $e^{-\chi_R(t)}$ describes the decay
of the Rabi contrast, while $\Delta_R(t)$ describes coherent quadrature mixing
generated by the full finite-time matrix response.

In the toggling frame, the averaged Bloch vector is
\begin{equation}
    \langle \mathbf r_I(t)\rangle = M_I(t)\mathbf r_0 ,
    \label{eq:app_MI_phase_def}
\end{equation}
where $M_I(t)$ is the Pauli-transfer matrix of the noise-induced error map and
$\mathbf r_0$ is the initial Bloch vector.  Returning to the laboratory rotating
frame gives
\begin{equation}
    \langle \mathbf r(t)\rangle
    =
    R_x(\Omega t)M_I(t)\mathbf r_0 .
    \label{eq:app_lab_map_phase}
\end{equation}
Let
\begin{equation}
    \mathbf v(t)
    =
    M_I(t)\mathbf r_0
    =
    \begin{pmatrix}
        v_x(t)\\
        v_y(t)\\
        v_z(t)
    \end{pmatrix}.
    \label{eq:app_v_definition}
\end{equation}
Using the convention
\begin{equation}
    R_x(\Omega t)
    =
    \begin{pmatrix}
        1 & 0 & 0\\
        0 & \cos\Omega t & -\sin\Omega t\\
        0 & \sin\Omega t & \cos\Omega t
    \end{pmatrix},
\end{equation}
the measured $z$ component is
\begin{equation}
    \langle r_z(t)\rangle
    =
    v_y(t)\sin(\Omega t)+v_z(t)\cos(\Omega t).
    \label{eq:app_rz_vy_vz}
\end{equation}
For an initial ground state, $\mathbf r_0=(0,0,-1)^T$, the ideal noiseless
result is $\langle r_z(t)\rangle=-\cos(\Omega t)$.

The general expression in Eq.~\eqref{eq:app_rz_vy_vz} can be rewritten as an
amplitude and phase-shifted Rabi oscillation,
\begin{equation}
    \langle r_z(t)\rangle
    =
    -\mathcal C_R^{\mathrm{mat}}(t)
    \cos\!\left[\Omega t+\Delta_R(t)\right],
    \label{eq:app_rz_amp_phase}
\end{equation}
where
\begin{equation}
    \mathcal C_R^{\mathrm{mat}}(t)
    =
    \sqrt{v_y^2(t)+v_z^2(t)}
    \label{eq:app_matrix_contrast}
\end{equation}
and
\begin{equation}
    \Delta_R(t)
    =
    \operatorname{atan2}
    \!\left[
        v_y(t),-v_z(t)
    \right].
    \label{eq:app_deltaR_general}
\end{equation}
Here $\operatorname{atan2}(y,x)$ denotes the quadrant-correct inverse tangent.
Equation~\eqref{eq:app_deltaR_general} is the matrix-response definition of the
phase correction.  The scalar envelope used in the main text corresponds to the
approximation
\begin{equation}
    \mathcal C_R^{\mathrm{mat}}(t)
    \simeq
    e^{-\chi_R(t)} .
\end{equation}

For the initial ground state, $\mathbf r_0=-\mathbf e_z$, Eq.~\eqref{eq:app_v_definition}
gives
\begin{equation}
    v_y(t)=-[M_I(t)]_{yz},
    \qquad
    v_z(t)=-[M_I(t)]_{zz}.
\end{equation}
Therefore,
\begin{equation}
    \mathcal C_R^{\mathrm{mat}}(t)
    =
    \sqrt{
        [M_I(t)]_{yz}^2+[M_I(t)]_{zz}^2
    },
    \label{eq:app_ground_state_contrast}
\end{equation}
and
\begin{equation}
    \Delta_R(t)
    =
    \operatorname{atan2}
    \!\left[
        -[M_I(t)]_{yz},
        [M_I(t)]_{zz}
    \right].
    \label{eq:app_deltaR_ground}
\end{equation}
In the weak-noise limit, $[M_I(t)]_{zz}\simeq 1$ and
$|[M_I(t)]_{yz}|\ll 1$, so
\begin{equation}
    \Delta_R(t)
    \simeq
    -[M_I(t)]_{yz}.
    \label{eq:app_deltaR_weak}
\end{equation}

A perturbative expression follows from the equation of motion
\begin{equation}
    \dot M_I(t)=K(t)M_I(t),
    \qquad
    M_I(0)=I_3 .
\end{equation}
To leading order in the noise strength,
\begin{equation}
    M_I(t)
    \simeq
    I_3+\int_0^t ds\,K(s).
\end{equation}
Thus
\begin{equation}
    [M_I(t)]_{yz}
    \simeq
    \int_0^t ds\,K_{yz}(s),
\end{equation}
and
\begin{equation}
    \Delta_R(t)
    \simeq
    -\int_0^t ds\,K_{yz}(s).
    \label{eq:app_deltaR_Kyz}
\end{equation}

Using
\begin{equation}
    K(t)
    =
    \int_0^t d\tau\,C(\tau)
    \left[
        \mathbf n(t-\tau)\mathbf n^T(t)
        -
        \cos(\Omega\tau)I_3
    \right],
\end{equation}
with
\begin{equation}
    \mathbf n(t)
    =
    \begin{pmatrix}
        0\\
        \sin(\Omega t)\\
        \cos(\Omega t)
    \end{pmatrix},
\end{equation}
the $yz$ component is
\begin{align}
    K_{yz}(s)
    &=
    \int_0^s d\tau\,C(\tau)
    n_z(s-\tau)n_y(s)
    \nonumber\\
    &=
    \int_0^s d\tau\,C(\tau)
    \cos[\Omega(s-\tau)]\sin(\Omega s).
    \label{eq:app_Kyz}
\end{align}
Equivalently,
\begin{align}
    K_{yz}(s)
    &=
    \frac{1}{2}
    \int_0^s d\tau\,C(\tau)
    \left\{
        \sin[2\Omega s-\Omega\tau]
        +
        \sin(\Omega\tau)
    \right\}.
    \label{eq:app_Kyz_split}
\end{align}
Substituting Eq.~\eqref{eq:app_Kyz} into Eq.~\eqref{eq:app_deltaR_Kyz} gives the
leading-order phase correction,
\begin{equation}
    \Delta_R(t)
    \simeq
    -
    \int_0^t ds
    \int_0^s d\tau\,C(\tau)
    \cos[\Omega(s-\tau)]\sin(\Omega s).
    \label{eq:app_deltaR_integral}
\end{equation}

Equation~\eqref{eq:app_deltaR_integral} shows that $\Delta_R(t)$ is generated
by the off-diagonal quadrature response of the full matrix map.  In the scalar
Rabi-envelope approximation, this coherent phase correction is usually small
compared with the contrast decay and can be absorbed into a fitted Rabi phase
or a small effective shift of the Rabi frequency.

\section{Rabi-frequency dependence of the phase-noise-limited single-qubit gate fidelity}
\label{app:fidelity_frequency_scan}

To examine how the servo-bump phase noise affects single-qubit gates across
operating conditions, we calculate the phase-noise-limited average fidelity of
a resonant $\pi$ pulse versus Rabi frequency. For each Rabi frequency
$f_{\mathrm{R}}$,
\begin{equation}
    \Omega = 2\pi f_{\mathrm{R}},
    \qquad
    t_{\pi} = \frac{\pi}{\Omega} = \frac{1}{2f_{\mathrm{R}}}.
    \label{eq:app_tpi_frequency}
\end{equation}

The measured delayed-self-heterodyne phase-noise spectrum is converted to the
detuning-noise power spectral density,
\begin{equation}
    S_{\beta}(2\pi f) = (2\pi f)^2 S_{\phi}(f),
    \label{eq:app_sbeta_conversion}
\end{equation}
with $\beta(t)=\dot{\phi}(t)$. A single spectral calibration factor, determined
from the feedforward-off Rabi-decay measurement, is used for all Rabi
frequencies and both feedforward conditions; no frequency-dependent
recalibration is performed.

\begin{figure*}[t]
    \centering
    \includegraphics[width=0.96\textwidth]
    {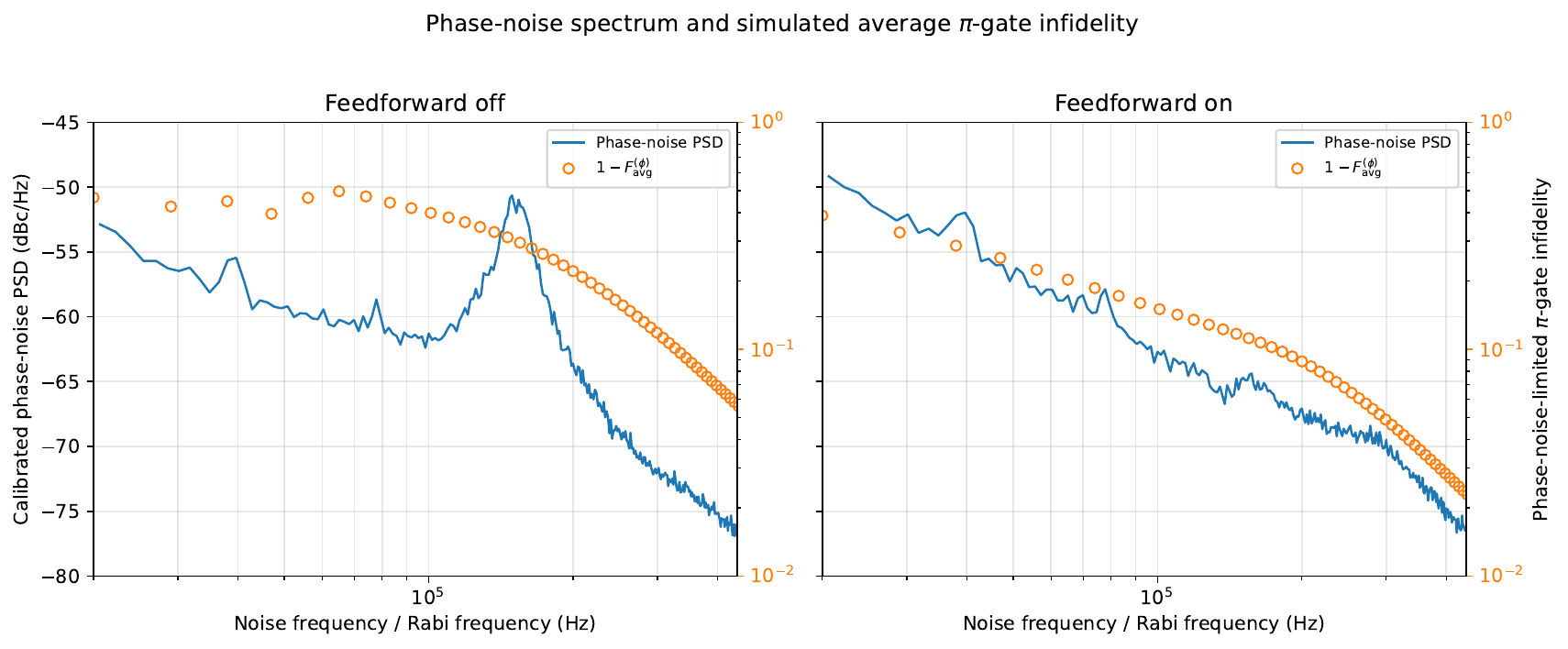}
    \caption{
    Phase-noise spectrum and simulated phase-noise-limited average $\pi$-gate
    infidelity versus frequency. Blue curves: calibrated phase-noise PSD (left
    axis, dBc/Hz). Orange open circles: $1-F_{\mathrm{avg}}^{(\phi)}$ for
    resonant $\pi$ pulses (logarithmic right axis), where at each marker the
    Rabi frequency equals that frequency and $t_{\pi}=1/(2f_{\mathrm{R}})$. The
    full measured spectrum, not just the PSD value at $f=f_{\mathrm{R}}$, enters
    each finite-time calculation. Left and right panels show PDH feedforward
    disabled and enabled. The Rabi frequency is scanned from $20$ to
    $450~\mathrm{kHz}$ with one common calibration factor.
    }
    \label{fig:fidelity_vs_frequency}
\end{figure*}

For each $f_{\mathrm{R}}$, the finite-time Pauli-transfer matrix $M_I(t_{\pi})$
is computed from the complete measured spectrum, giving the average gate
fidelity
\begin{equation}
    F_{\mathrm{avg}}^{(\phi)}(t_{\pi})
    = \frac{3+\operatorname{Tr}[M_I(t_{\pi})]}{6},
    \label{eq:app_frequency_fidelity}
\end{equation}
and the infidelity shown in Fig.~\ref{fig:fidelity_vs_frequency},
$\epsilon_{\mathrm{avg}}^{(\phi)} = 1-F_{\mathrm{avg}}^{(\phi)}$.

The feedforward-off infidelity peaks near $f_{\mathrm{R}}\simeq63~\mathrm{kHz}$
($1-F_{\mathrm{avg}}^{(\phi)}\simeq0.498$), below the servo-bump center. Because
a short $\pi$ pulse probes a finite band rather than a single Fourier frequency,
lowering $f_{\mathrm{R}}$ lengthens the pulse and reshapes the finite-time
spectral response; the integrated overlap between gate response and phase noise
can thus peak below the PSD maximum, consistent with
Ref.~\cite{nakav2023effect}.

Enabling feedforward reduces the infidelity throughout the region around the
servo bump. The largest feedforward-on value shown occurs at the scan boundary,
$f_{\mathrm{R}}=20~\mathrm{kHz}$ ($1-F_{\mathrm{avg}}^{(\phi)}\simeq0.388$), and
should not be read as a resolved internal maximum; calculations at very low
Rabi frequencies also grow sensitive to phase noise below the lowest measured
DSHI frequency.

These are model-calibrated estimates of the laser-phase-noise contribution
only. They exclude SPAM, intensity noise, magnetic-field fluctuations,
pulse-area errors, and other technical noise.

\section{Log-linear dependence of $T_2$ on phase-noise suppression in dB}

Within the filter–function picture, the dephasing rate $\Gamma = T_2^{-1}$ is proportional to the laser phase-noise power spectral density (PSD) at the Rabi frequency,
\[
\Gamma \propto S_\phi(f = \Omega / 2\pi).
\]
When the servo-bump noise is suppressed by $\Delta_{\mathrm{dB}}$ (in dB), the PSD scales as
\[
S_\phi' = S_\phi \times 10^{-\Delta_{\mathrm{dB}}/10}.
\]
Accordingly, the dephasing rate becomes
\[
\Gamma' = \Gamma_0 \times 10^{-\Delta_{\mathrm{dB}}/10},
\]
which implies the coherence time follows the log–linear relation
\[
T_2(\Delta_{\mathrm{dB}}) = T_{2,0} \times 10^{\Delta_{\mathrm{dB}}/10},
\]
where $T_{2,0}$ denotes the coherence time at zero additional suppression.

Taking the base-10 logarithm yields a linear expression,
\[
\log_{10} T_2 = \log_{10} T_{2,0} + \frac{1}{10} \, \Delta_{\mathrm{dB}},
\]
indicating that a linear fit of $\log_{10}(T_2)$ versus $\Delta_{\mathrm{dB}}$ is the most appropriate model across the entire dynamic range.


\section{Thulium-Doped Fiber Amplifier (TDFA)}

Thulium-doped fiber amplifiers (TDFAs) provide high-power amplification in the 1.6–2.0~µm band, enabling narrow-linewidth sources for applications such as optical qubits, spectroscopy, and telecom systems. The gain medium is a silica fiber doped with Tm$^{3+}$ ions, which amplify seed light via stimulated emission on the $^3$F$_4 \rightarrow ^3$H$_6$ transition once population inversion is achieved.

Two common pumping schemes are used: (i) \emph{off-band pumping} at 790~nm via the $^3$H$_6 \rightarrow ^3$H$_4$ transition followed by nonradiative decay to $^3$F$_4$, and (ii) \emph{in-band pumping} at 1550~nm, which directly populates $^3$F$_4$ and offers higher efficiency and lower heat load. In our system, the TDFA is in-band pumped with a 1550~nm laser and seeded by a 1762~nm ultra-stable cavity-locked ECDL.

When the seed intensity is sufficiently high, amplified spontaneous emission (ASE) is suppressed, and the amplified linewidth remains close to that of the seed. We characterized the amplified laser using the delayed self-heterodyne interferometry (DSHI) method with a 5~km delay fiber and 7~MHz EOM shift. Voigt-profile fitting yielded a combined Lorentzian linewidth of $320 \pm 30$~Hz, corresponding to a single-laser linewidth of $160 \pm 15$~Hz.

Independent verification via high-resolution spectroscopy on the $\mathrm{S}_{1/2}(m_j=-1/2) \leftrightarrow \mathrm{D}_{5/2}(m_j=-1/2)$ transition of a single $^{138}$Ba$^+$ ion produced consistent results ($156 \pm 16$~Hz), confirming negligible linewidth broadening. Rabi oscillations revealed slightly shorter coherence time for the TDFA beam ($T_2 = 97~\mu$s) compared to the master laser ($T_2 = 125~\mu$s), attributed to higher low-frequency intensity noise rather than phase noise. Power-broadening measurements further supported an intrinsic linewidth below 200~Hz.

In summary, the in-band pumped TDFA delivers $\sim$150~mW output at 1762~nm while preserving the seed’s sub-200~Hz linewidth across a wide tuning range. The main limitation arises from intensity noise, which can be mitigated by active stabilization. This confirms TDFAs as suitable amplifiers for high-fidelity trapped-ion quantum gates and narrow-linewidth optical systems~\cite{ahmadi2024scalable,walasik20231760}.

\bibliographystyle{apsrev4-2}
\bibliography{references}

\end{document}